\documentclass[pra,aps,showpacs,superscriptaddress,twocolumn,longbibliography,amsmath,amssymb]{revtex4-1}

\usepackage{graphicx}
\usepackage{bm}
\usepackage[normalem]{ulem} 							
\usepackage[usenames, dvipsnames]{color} 				
\usepackage{subfigure}
\usepackage[english]{babel}
\usepackage{nicefrac} 

\usepackage[breaklinks=true]{hyperref}
\hypersetup{
  colorlinks   = true, 
  urlcolor     = black, 
  linkcolor    = black, 
  citecolor   =  black 
}

\def\etal{\textit{et~al.\ }}
\def\insitu{\textit{in situ\ }}

\newcommand{\ket}[1]{|#1\rangle}
\newcommand{\bra}[1]{\langle #1|}

\newcommand{\op}[1]{\hat{#1}}

\newcommand{\comm}[2]{\left[ #1, #2 \right]}
\newcommand{\diss}[1]{\mathcal{D} [#1]}

\newcommand{\hc}{^{\dagger}}						
\newcommand{\ee}{\mathrm{e}}						
\newcommand{\ii}{\mathrm{i}}			             			
\newcommand{\nn}{\nonumber}							

\DeclareRobustCommand{\SkipTocEntry}[4]{}

\begin{document}

\title{Towards understanding two-level-systems in amorphous solids - Insights from quantum circuits} 

\author{Clemens M\"uller}
\email{clm@zurich.ibm.com}
\affiliation{IBM Research Zurich, 8803 R\"uschlikon, Switzerland}
\affiliation{Institute for Theoretical Physics, ETH Z\"urich, 8093 Z\"urich, Switzerland}
\affiliation{ARC Centre of Excellence for Engineered Quantum Systems, School of Mathematics and Physics, The University of Queensland, Brisbane, Queensland 4072, Australia}

\author{Jared H. Cole}
\email{jared.cole@rmit.edu.au}
\affiliation{Chemical and Quantum Physics, School of Science, RMIT University, Melbourne, Victoria 3001, Australia}

\author{J\"urgen Lisenfeld}
\email{juergen.lisenfeld@kit.edu}
\affiliation{Physikalisches Institut, Karlsruhe Institute of Technology, 76131 Karlsruhe, Germany}



\date{\today}
             
\begin{abstract}
	Amorphous solids show surprisingly universal behaviour at low temperatures. The prevailing wisdom is that this can be explained by the existence of two-state defects within the material. 
	The so-called \emph{standard tunneling model} has become the established framework to explain these results, yet it still leaves the central question essentially unanswered - what are these two-level defects?  
	This question has recently taken on a new urgency with the rise of superconducting circuits in quantum computing, circuit quantum electrodynamics, magnetometry, electrometry and metrology.  
	Superconducting circuits made from aluminium or niobium are fundamentally limited by losses due to two-level defects within the amorphous oxide layers encasing them.  
	On the other hand, these circuits also provide a novel and effective method for studying the very defects which limit their operation. 
	We can now go beyond ensemble measurements and probe individual defects - observing the quantum nature of their dynamics and studying their formation, 
	their behaviour as a function of applied field, strain, temperature and other properties.
	This article reviews the plethora of recent experimental results in this area and discusses the various theoretical models which have been used to describe the observations. 
	In doing so, it summarises the current approaches to solving this fundamentally important problem in solid-state physics.
\end{abstract}

\maketitle

\tableofcontents{}



\section{Introduction\label{sec:Intro}}

\textbf{Two-Level-Systems in amorphous materials - }
The properties of amorphous or glassy material are dominated by the absence of long-range order in the atomic lattice. 
Even though they have been studied for decades, there is still a surprisingly large amount that is not understood about such materials. 
Despite the randomness of the atomic arrangements, and even independent of their chemical composition, 
most amorphous solids display surprising similarities in their properties at temperatures below a few Kelvin. 
This \emph{universality} is typically explained via the so-called standard tunneling model (STM), whose basic principle is that the low-temperature behaviour of glassy systems is dominated 
by the presence of two-level defects (TLS) within the material. 
These defects are in general not due to impurities inside the materials, but rather emerge from the deviations away from crystalline order which characterise the amorphous state. 
Due to their low energy, such two-level defects are typically saturated at high temperatures. 
However as the material is cooled, these additional degrees of freedom become available and can dominate the low temperature properties.

In general, the STM is exceptionally good at describing the low-temperature behaviour of most amorphous materials~\cite{Enss:2005, Esquinazi:2013}.
As this approach treats the defects at a phenomenological level, it can be used across many different systems. 
However, it leaves one fundamentally important question unanswered - what is the underlying microscopic nature of the two-level defects?

\textbf{Two-Level-Systems in Quantum Devices - }
Recently, TLS have attracted substantial renewed interest because they are seen as a major source of noise and decoherence in superconducting quantum devices. 
This includes superconducting quantum bits (qubits), which are circuits with resonance frequencies in the microwave range, tailored from microstructured inductors, 
capacitors, and Josephson tunnel junctions (JJs)~\cite{Clarke:N:2008, Devoret2004, Devoret:Science:2013}. 
To operate such a circuit as a qubit, it is necessary to have two long-living eigenstates which are used as logical states $\ket{0}$ and $\ket{1}$, 
and between which transitions can be driven to realize logical quantum gates. 
Since such circuits in general have more than two excited states, they need to be sufficiently anharmonic so that all transition frequencies are unique and can be separately addressed. 
This is realized by incorporating Josephson junctions, which can be modelled as nonlinear inductors whose value is tuned via a bias current or an applied magnetic flux. 
\begin{figure}[b]
	\includegraphics{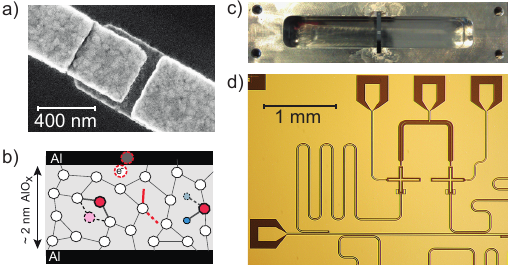}
	\caption{a) SEM photograph of a Josephson junction (JJ) made with the aluminium shadow evaporation technique. 
		b) Sketch of a JJ formed by two superconducting Al electrodes that are separated by a thin ($\approx 2-3$ nm) layer of amorphous AlO$_x$ dielectric, 
		here illustrated as hosting TLS formed by tunneling atoms, dangling bonds, and trapped charges.	
		c) Photograph of a 3D-Transmon qubit, showing the opened 3-dimensional cavity and a qubit chip in its centre, similar to~\textcite{Paik:PRL:2011}. 
		d) Planar Transmon qubits, consisting of cross-shaped capacitor electrodes shunted by JJs. 
		Meander structures are resonators coupled to a transmission line for qubit readout, similar to~\textcite{Barends:PRL:2013}.
	}
	\label{fig:QubitsAndTLS}
\end{figure}
Since the first observation of coherent quantum dynamics in a Cooper pair box in the year 1999~\cite{Nakamura:N:1999}, 
superconducting qubits have evolved into one of the leading contenders for the realization of large-scale quantum computing~\cite{Barends:Nature:2014, Mohseni:Nature:2017}.

However, loss and fluctuations due to parasitic coupling to TLS present a significant source of decoherence {and parameter fluctuations} for superconducting qubits~\cite{Steffen:Sust:2017}. 
At present, quantum circuits are typically fabricated from superconducting aluminium because it allows the formation of high-quality JJs 
using the well-established techniques of double angle shadow evaporation~\cite{Niemeyer:APL:1976,Dolan:APL:1977,Dolan:PhysicaB:1988} 
(see Fig.~\ref{fig:QubitsAndTLS}a) for an example). 
Once the sample chips are exposed to air, an amorphous oxide layer will grow on any exposed aluminium structure, 
which is characterized by a large dielectric loss that in turn is attributed to high TLS densities in the amorphous material~\cite{Martinis:arxiv:2014}. 
Moreover, the insulating tunnel barrier of JJs is itself also made from amorphous aluminium oxide and thought to host TLS. 
Figure~\ref{fig:QubitsAndTLS}b) illustrates some models of TLS formation in Josephson junctions. 

When the two states of a TLS are associated with the displacement of a charge, they possess an electric dipole moment which couples them to the oscillating electric field present in capacitive circuit components. 
For TLS residing in the typically few nm-thin tunnel barrier of JJs, where the electric field strength can reach several 100 V/m, this coupling may become particularly strong. 
While such strongly coupled TLS are especially detrimental to qubit operation, their coherent interaction with the circuit dynamics provides a pathway 
towards direct manipulation and readout of the state of the microscopic defects using their macroscopic host device.

When we consider TLS in quantum devices, there is a useful classification to bear in mind which relates to whether the TLS internal dynamics on experimental timescales 
is dominated by incoherent or 
processes or not. 
Each individual TLS is coupled to an environment at ambient temperature $T$ which might for example consist of phonon modes, other TLS in their vicinity 
or quasiparticles formed from residual non-superconducting charge carriers. This coupling leads to incoherent transitions between the TLS eigenstates (dissipation and excitation) 
as well as random fluctuations in their energy (dephasing). 

\textbf{Fluctuators} can be defined as TLS who are in strong contact with their own environment and incoherently flip between two states on typical experimental timescales. 
These incoherent state transitions are due to a combination of quantum tunnelling through the barrier and decoherence due to the coupling to their environment. 
The additional option of thermal activation, i.e. when the thermal energy $k_{B}T$ of the environment is larger than the height of the barrier between the two wells,
 is typically not relevant for the tunneling two-level systems considered here. 
In low-temperature electronics the fluctuators can couple to their host circuit, and the resulting fluctuations will manifest themselves as quasi-classical random variations in circuit parameters. In addition, through defect-defect coupling they may also modify the behaviour of other TLS in the ensemble.
Since very fast fluctuations will average out over experimental timescales, fluctuators will typically modify the dynamics of quantum devices via contributions to the low-frequency environmental noise spectrum~\cite{Dutta:RMP:1981}, 
which in turn is mostly responsible for loss of phase coherence in these devices~\cite{Paladino:RMP:2014}.

\textbf{Coherent TLS}, on the other hand, are those where the coupling between the TLS and their environment is weak enough that they can remain in one of their eigenstates, or they can even be placed in a coherent superposition of states, over the timescale of an experiment.
Typically, coherent TLS will have energy splittings that are larger than the thermal energy of their environment, $E > k_{B} T$, 
such that incoherent excitations into their excited state are suppressed and their equilibrium steady-state will be their ground state. 
Such coherent TLS can even reach the strong coupling regime with their host circuit or each other, which is characterized by a coupling strength that exceeds the decoherence rates of both the TLS and its hosting device. 
This strong coupling results in modifications of the energy level structure and quantum dynamics of the hosting device which can be directly observed, 
for example as anti-crossings in qubit spectroscopy~\cite{Simmonds:PRL:2004} or coherent beating in population dynamics~\cite{Neeley:NaturePhysics:2008}. 
Coherent TLS in their groundstate are also able to resonantly absorb energy from their host quantum circuits and dissipate it into their own environment~\cite{Shnirman:PRL:2005, Barends:PRL:2013}.
The distinction between fluctuators and coherent TLS introduced here is not fundamental and they are in fact thought to be formed by the same physical entities. However it provides a useful distinction when considering the possible dynamical effects that can arise from an individual TLS or the effect of an entire ensemble~\cite{Wold:PRB:2012}.

We also note that the classification given above is distinct from the term ``incoherent TLS'', which in the glass physics community is taken to imply a TLS for which incoherent, environment induced processes dominate over all other energy scales, or in other words a TLS whose dynamics is overdamped. 

\textbf{State of the art - }
The impressive enhancement of coherence times that was seen in experiments with superconducting qubits during the last decade~\cite{Steffen:Physics:2011, Devoret:Science:2013}, 
was to a large extend achieved by reducing the coupling to TLS {(and other decoherence sources)} with improved circuit designs, rather than by lowering the TLS densities by improving materials and fabrication procedures. 
The best performing qubits today employ small tunnel junctions to reduce the number of strongly coupled TLS, and employ circuit layouts in which the electric field concentration is reduced at lossy interfaces, 
e.g., by increasing the distances between capacitor electrodes. This is clearly demonstrated in so-called three-dimensional ``Transmon'' qubits~\cite{Paik:PRL:2011, Rigetti:PRB:2012}, 
which feature large capacitor plate separation and are placed into cavity resonators having large volumes of a few $cm^3$, 
such that the electric field strength is significantly lower than in other designs (see Fig.~\ref{fig:QubitsAndTLS}c) for an example).
Careful revision of clean-room recipes has been identified as a second necessity in order to avoid the formation of TLS during sample fabrication. 

The growing understanding of how to evade the TLS problem has brought superconducting circuits in short time to the verge of becoming scaled up to integrated quantum processors. 
Nevertheless, dielectric loss remains responsible for the major part of energy relaxation in state-of-the art qubits~\cite{Wang:APL:2015}, 
and coupling to even sparse TLS baths causes relaxation~\cite{Mueller:PRB:2015} and dephasing~\cite{Faoro:PRB:2015}.
These issues will gain in importance once circuits comprise more than a handful of prototype qubits, 
and thus must be urgently addressed to ensure continuation on the path towards a solid-state quantum computer.

On the other hand is it just this sensitivity to even single defects that makes superconducting circuits ideal tools for the study of TLS and material dissipation mechanisms in the quantum 
(i.e.\ single-photon and low-temperature) regime. 
The possibility of using superconducting qubits to probe individual defects has fundamentally changed both the questions that \emph{can be} asked and that \emph{need to be} asked.  
These circuits allow not just the dissipative dynamics of a defect to be resolved directly, but even enable one to measure and manipulate the quantum states of coherent TLS.  
This has opened doors to new tests and studies of the nature of individual defects and therefore the ensemble as a whole.

\textbf{Outline - }
In this review, we focus on TLS which affect novel superconducting quantum circuits such as qubits and resonators. 
With qubits, one is able to access and control the quantum states of individual TLS, enabling novel studies of their mutual interactions and decoherence mechanisms.
Microwave resonators on the other hand are effective tools to characterize loss from defects at layer interfaces and to quickly validate fabrication processes. 
They are also a necessary part of the leading solid-state architecture for quantum computation, circuit-QED~\cite{Wallraff:nature:2004, Blais:PRA:2004}, 
where qubits are coupled to resonators to improve interactions and readout. 
We further review the existing theoretical models for the origin of TLS and how these can be reconciled with existing and future experiments.

The review begins with a short introduction of the standard tunneling model, and how ensembles of TLS are a source of low- and high-frequency noise for superconducting circuits.
We continue in Sec.~\ref{Sec:Models} with an overview of the plethora of proposed microscopic models for the origin of TLS. 
The following Section~\ref{Sec:Interactions} gives a brief discussion of the basic physical mechanisms by which TLS interact with superconducting circuits and their environment
and how these interactions can be utilised to draw conclusions on their microscopic origin. 
Section~\ref{sec:TLSinqubits} presents an overview of experiments on qubits such as spectroscopy, by which the presence of coherent coupling to individual TLS was first revealed. 
It also reviews measurements of the coherent time evolution of TLS quantum states, including quantum state swapping, creation of TLS entanglement, 
measurements of TLS decoherence, and studies of mutual TLS-TLS interaction.
Most of the knowledge about materials and fabrication steps which give rise to TLS formation has been obtained from experiments on superconducting resonators, 
which we describe in Sec.~\ref{sec:resonators}. 
The following Sec.~\ref{sec:othersystems} gives a very brief summary of other experimental architectures where TLS are believed to be of relevance, such as nano-mechanical resonators.
Finally, we give an overview of recent progress and results from fabrication of superconducting circuits in Sec.~\ref{sec:fabrication}, 
which also includes a review of ongoing efforts to fabricate Josephson junction with crystalline tunnel barriers.
For additional recent reviews on the importance of materials in superconducting quantum bits we refer the reader also to~\textcite{McDermott:IEEE:2009, Oliver:2013}. 
We end with a short summary and outlook, aiming to provide a perspective for future experimental and theoretical efforts to determine the microscopic origins of TLS.


\section{Background\label{sec:Background}}

\subsection{Standard tunneling model}
The concept that a single atom can tunnel between energetically equivalent local potential wells was understood even at the birth of quantum mechanics~\cite{Hund:ZPhys:1927,Cleeton:PR:1934}. 
Soon after, \textcite{Pauling:PR:1930} discussed the implications for the oscillatory and rotational motion of atoms in molecules and crystals. 
However, more recently there has been a vast array of different models proposed for the physical origin of TLS.  
The details of many of these models will be discussed in section~\ref{Sec:Models} but they all share some fundamental properties which form the basic components of the standard tunneling model (STM)~\cite{Philips:JLowTemp:1972,Anderson:PhilMag:1972,Philips:1981}.  
In this section, we briefly summarise the key aspects of the standard tunneling model which will be important for the later discussions.  
A more in-depth overview of the STM and its supporting experiments is given in~\textcite{Esquinazi:2013, Enss:2005,Wuerger:1997,Lubchenko:AdvancesChemPhys2007}. 

\begin{figure}[tbh!]
	\includegraphics{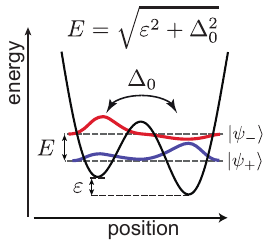} 
	\caption{Double-well potential modelling a TLS. The energy difference $E$ between the TLS eigenstates $\ket{\psi_+}$ and $\ket{\psi_-}$ is determined by the asymmetry energy $\varepsilon$ and the inter-well tunneling rate $\Delta_0$.}	
	\label{fig:DWP}
\end{figure}

The standard tunneling model makes the following key assumptions
\begin{itemize}
	\item The TLS can exist in one of two energetically similar configurations.
	\item These configurations are modelled as two minima in a double-well potential which are separated by a barrier.
	\item At sufficiently low temperatures, thermal activation over the barrier is suppressed and the dynamics are governed by quantum tunneling through the barrier.
	\item In general, the system couples to applied electric or strain fields in such a way that transitions can be driven between the states.
	\item Due to the random atomic arrangements, an ensemble of TLS is characterized by a wide distribution of potential barrier heights and thus spans a large range of switching rates and eigenenergies.
\end{itemize}
A common visualisation of a TLS in the STM is given in Fig.~\ref{fig:DWP} where a particle can sit in one of two parabolic potential wells.  
The energy asymmetry of the ground-state wavefunctions of the two wells is labelled $\varepsilon$. This asymmetry can be due to differences in width or shape of the two wells, or of the classical minimum energy of the wells (as depicted).  
The energy associated with the process of tunneling through the barrier separating the two wells is $\Delta_0$.  

The effective Hamiltonian for this situation has the form
\begin{equation}\label{eq:HTLS}
	H_{\rm{TLS}} = \frac12 \left( \begin{array}{cc} \varepsilon & \Delta_0 \\ \Delta_0 & -\varepsilon \end{array} \right) = \frac12 \varepsilon \sigma_{z}^{(p)} + \frac12 \Delta_{0} \sigma_{x}^{(p)}, 
\end{equation}
with the Pauli matrices in the position basis $\sigma_{z}^{(p)} = \ket R \bra R - \ket L \bra L$ and $\sigma_{x}^{(p)} = \ket R \bra L + \ket L \bra R$.
Note that for clarity, our notation here differs from the usual STM literature where typically the asymmetry energy $\varepsilon$ is labelled $\Delta$. 
Due to the tunneling between the wells, the two lowest eigenstates in the left ($\ket{L}$) and right ($\ket{R}$) well hybridise and form the eigenstates 
\begin{eqnarray}
	\ket{\psi_+} = \sin\left( \frac{\theta}{2} \right) \ket{L} + \cos\left( \frac{\theta}{2} \right) \ket{R} \,,\\
	\ket{\psi_-} = \cos\left( \frac{\theta}{2} \right) \ket{L} - \sin\left( \frac{\theta}{2} \right) \ket{R} \,,
\end{eqnarray}
where the mixing angle $\theta$ is defined via $\tan \theta = \Delta_0/\varepsilon$.
We can then rewrite the Hamiltonian in the basis of eigenstates as
\begin{align}
	H_{\rm{TLS}} = \frac12 E \sigma_{z} \,,
	\label{eq:HTLSEigen}
\end{align}
with the energy difference between the eigenstates 
\begin{align}
E = E_{+} - E_{-} = \sqrt{\varepsilon^2 + \Delta_0^2}.
\label{eq:tls_energy}
\end{align}
In the limit $|\varepsilon|\gg\Delta_0$, the eigenstates are well described by the left and right well states.  
However for $|\varepsilon|\approx0$, the eigenstates are a superposition of the two well states.

Using Wentzel-Kramers-Brillouin (WKB) theory, we can estimate the value of $\Delta_0$ in terms of the barrier height $V$, the spacing between the wells $d$ and the effective mass of the particle $m$~\cite{Esquinazi:2013}, giving
\begin{equation}
	\Delta_0 = \hbar \omega_0 e^{-\lambda} \,,
\end{equation}
where
\begin{equation}
	\lambda = \sqrt{\frac{2 m V}{\hbar^2}}d.
\end{equation}
The scale factor $\omega_0$ depends on the exact functional form of the potential, see~\textcite{Philips:1981} for examples.
However, this detail is typically unimportant for the overall behaviour of the system given the exponential dependence on $\lambda$. 

%
	
In contrast to disorder or impurity defects in crystals which can also display TLS behaviour, see~\cite{Enss:2005} for a detailed discussion), 
the TLS parameters in amorphous solids vary from defect to defect due to the randomised nature of the local atomic configurations.  
The STM assumes that the asymmetry $\varepsilon$ and tunneling parameter $\lambda$ are independent and uniformly distributed, $P(\varepsilon, \lambda)d\varepsilon d\lambda = P_0 d\varepsilon d\lambda$, 
where $P_0$ is a constant. Re-expressing in terms of $E$ and $\Delta_0$,	
\begin{equation}
	P(E,\Delta_0) d\Delta_0 dE = P_0 \frac{E}{\Delta_0\sqrt{E^2-\Delta_0^2}} d\Delta_0 dE \,,
\end{equation} 
and integrating over $\Delta_0$ gives the density of states,
\begin{equation}
	D(E)=\int_{\Delta_0^{\rm{min}}}^E P(E,\Delta_0) d\Delta_0 = P_0 \ln \frac{2E}{\Delta_0^{\rm{min}}} \approx D_0 \,,
	\label{eq:TLSDensity}
\end{equation}
which is typically approximately constant over the energy ranges of interest.
Here $\Delta_0^{\rm{min}}$ refers to the minimum tunneling energy, below which the particle can no longer tunnel between the wells and the TLS is essentially just two local minima in a classical sense. 
Typically, the observation timescale of an experiment 
(which can vary from $10^{-6}$ to $10^{3}$ seconds) sets a minimum energy scale of interest. 
In amorphous glasses in general, the distribution of $\Delta_0$ extends far below any energy set by the experimental timescale (i.e.\ $\Delta_{0}^{\rm{min}}\ll E$) which supports the constant density of states approximation. 
Using this density of states and computing the specific heat of a material containing TLS defects, one finds that $C_V\propto T$, as opposed to the usual $T^3$ obtained from Debye theory. 
This modified temperature dependence is one of several key predictions of the STM for the low-temperature behaviour of amorphous solids.

Although the STM typically assumes independent TLS, including the effects of interactions between TLS leads to corrections to the low-temperature response of glasses due to the formation of collective states. 
Although such extensions to the STM have been studied in depth in glasses both experimentally and theoretically, 
recent work with superconducting circuits has provided far more direct evidence for TLS-TLS interactions. 
We elaborate more on such models in section~\ref{Sec:Models:Emergent} where we provide examples of collective models for TLS origins, 
and section~\ref{Sec:Interactions}, where we discuss direct probes of TLS-TLS interactions. 


\subsection{TLS as a source of low-frequency noise\label{Sec:Noise}}

The STM and its extensions are often used as models of electrical noise in general, especially in low-temperature electronic devices. 
Typically, this noise is thought to arise as an ensemble effect from a large number of TLS with some distribution of energies and tunnelling rates. 
Here, we briefly touch on noise in electrical circuits in general, so as to introduce the concepts and notation relevant to our later discussion.
We will focus first on the low-frequency noise spectrum, and discuss the high-frequency noise components relevant to energy dissipation in the following section.

When characterizing a fluctuating quantity $X(t)$ (e.g. an applied voltage), one usually defines the spectral function of this quantity through the Fourier transformation of its two-time correlation function
\begin{align}
	S_{X}(\omega) = \int_{-\infty}^{\infty} \ee^{-\ii \omega t} \left\langle X(t) X(0) \right\rangle .
\end{align}
At low frequencies and temperatures, the dominant noise source in electrical solid-state devices typically has the functional form
\begin{align}
	S_{X} (\omega) \propto \omega^\alpha \,,
\end{align}
which, when $\alpha \sim -1$, is the (in)famous `1/f noise'. 

This contribution to the noise has been studied for decades and although it is ubiquitous across many different types of devices and experiments, at this stage it is still poorly understood. 
Recently, due to its detrimental effects on coherent superconducting circuits, the need to understand the microscopic origin of this noise has been given a new urgency 
(see~\onlinecite{Weissman:RMP:1988} and~\onlinecite{Paladino:RMP:2014} for more specialised and in-depth reviews of 1/f noise in condensed matter physics and qubits). 
Qubit coherence depends on both the noise spectral function at the resonance frequency of the circuit \emph{and} at or close to zero frequency. 
It is common that coherence times are limited by the $1/f$ noise power (see section~\ref{sec:TLSinqubits} for more details) making the understanding of its origins of paramount importance.

The random nature and distribution of relaxation times characteristic of the STM already suggests that an ensemble of TLS might provide a viable model for the $1/f$ noise. 
Although~\textcite{Dutta:RMP:1981} showed several decades ago that a distribution of TLS switching rates $P(\gamma)\sim 1/\gamma$ together with the constant distribution in energy from the STM results in the required spectral characteristics of the noise,  
the link between a specific case of a TLS ensemble and $1/f$ noise in the same device has so far been difficult to prove definitively.



\subsection{TLS as source of high-frequency noise}

{At high frequencies ($\hbar \omega \gg k_B T$) electrical noise typically scales with $\alpha>0$, where $\alpha \sim 1$ is referred to as ``ohmic'' noise. 
This is the Johnson-Nyquist limit of noise and is traditionally explained as originating from a collection of linear harmonic oscillators, i.e. the modes of the waveguides and cables in the experiments and control electronics. 
However, ~\textcite{Shnirman:PRL:2005} showed that assuming a distribution of STM parameters, $P(\varepsilon, \Delta_{0}) = (\varepsilon/\Delta_{0})^{s}$ with $-1\leq s \leq 1$, 
leads to a linear distribution of TLS energies $E$ which results in a noise spectral density $S\propto 1/f$ for $hf \ll k_B T$ while at the same time giving the standard `Ohmic' signature ($S\propto f$) at high frequencies $hf \gg k_B T$. 
Apart from ensembles, even sparse distributions of TLS with eigenenergies comparable to the circuit frequencies can lead to dissipation, as they can accept energy from the circuit and dissipate it into their own environment~\cite{Muller:PRB:2009}.
This will typically manifest in a noise spectral function with pronounced peaks at certain frequencies, as observed commonly in superconducting qubits~\cite{Kakuyanagi:PRL:2007, Ithier:PRB:2005, Barends:PRL:2013, Paik:PRL:2011}.
}
{The cross-over between low and high-frequency noise regimes has also been studied experimentally (see for example~\onlinecite{Astafiev:PRL:2004, Quintana:PRL:2017})
and shown to appear at energies corresponding to the experimental temperature, consistent with the theoretical model of~\textcite{Shnirman:PRL:2005}
}
\\

{
\textbf{Summary} - 
The utility of the STM introduced here is that the majority of observed phenomena can be described with only a few parameters, 
namely $\varepsilon$, $\Delta_0$, the distributions of these parameters for a TLS ensemble and the TLS' dipole response. 
However, the recent work on TLS physics in superconducting circuits has focused on finding ways to not just understand, but actually remove TLS within the metal oxide surfaces and junctions. 
Therefore, a phenomenological model is insufficient, we need to know \emph{what a TLS is}, not just how it behaves. 
Although the double-well potential model illustrated in Fig.~\ref{fig:DWP} provides an intuitive picture of how the TLS parameters might come about, 
remaining `unknowns' in the problem are microscopic parameters such as particle mass, the charge of the tunnelling entity, and the size and form of the TLS potential. 
}



\section{Proposed microscopic models for the origin of TLS\label{Sec:Models}}

To date, all reported experiments are broadly consistent with the assumption that the TLS in superconducting circuits are equivalent to those known to exist 
in amorphous dielectrics such as glasses~\cite{Phillips:1987,Anderson:PhilMag:1972}. 
While these TLS in glasses have been studied intensively during the last 40 years, their microscopic nature remains elusive~\cite{Leggett:ACS:2013,Yu:JLTP:2004}.

Many different proposals exist to explain the origin of TLS in amorphous oxides within superconducting circuits, the major categories of which are summarised in this section. 
Due to the random nature of the oxide structure, there is no clear reason to expect unique spectral signatures of a particular TLS type - in contrast to typical defects in crystals which reside in a more well-defined environment.
However, given the additional information obtained from more recent experiments on strongly coupled (coherent) TLS, there is hope that an accurate comparison to theoretical models is possible. 
These prospects have encouraged several groups to investigate detailed computational models of TLS in amorphous material and then attempt to estimate the values of the resulting TLS parameters 
and their response to pulse sequences, applied strain, electric fields or temperature variation.

\subsection{Tunneling atoms}
\begin{figure}[bth!]
		\includegraphics{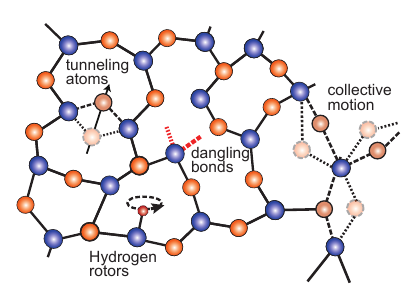} 
		\caption{Illustration of some example mechanisms of TLS formation in an amorphous material: tunneling of single atoms and collective motion of small atomic groups, dangling bonds, and Hydrogen defects.}
		\label{fig:TLSModels}
	\end{figure}

	One of the most physically appealing models is to assume that the two-level system is formed by the literal movement of an atom or small group of atoms between two potential minima. 
	An illustration of such systems in an amorphous material is shown in Fig.~\ref{fig:TLSModels}, depicting the motional degrees of freedom of atoms, dangling electronic bonds, and hydrogen atoms.
	
	Although this has traditionally been the physical picture that is most often quoted for the STM, it still leaves open the question of which atom or degree of freedom is actually `moving'.
	
	For accurate computational simulations of these models one has to grapple with several problems from a quantum chemistry point of view.  
	The energy splittings of typical strongly coupled TLS observed in experiments are 0.5~-~10 GHz, (ie.~neV) which is far too small for the majority of ab-initio methods.  
	In addition, the typical length scales of interest when studying metallic-oxides are 1-100 nm, involving 100s, 1000s or even more atoms. 
	Recent approaches to this problem have focussed on the specific problem of aluminium-oxide tunnel junctions, most relevant for TLS in superconducting circuits, contrasting with the more generic glass studies of the 80s and 90s.
	One is then able to use first principles methods, or empirical and effective potentials which have been optimised to correctly model the aluminium-oxide bonding configuration within the oxide.  
		
	One of the original suggestions for the origin of TLS in AlO$_\mathrm{x}$ and SiO$_2$ amorphous films observed with qubits are OH bonds or defects~\cite{Martinis:PRL:2005, Shalibo:PRL:2010}. 
	\textcite{Gordon:SciRep:2014} discussed the possibility that hydrogen interstitials within the Al$_2$O$_3$ lattice or at the surface could form suitable two-level defects. 
	They considered the stability of the various charge states of the hydrogen interstitial and the resulting structural geometries, and
	then solved the Schr\"odinger equation for the hydrogen atom within the potential formed by the surrounding Al$_2$O$_3$ structure - obtained directly from ab-initio methods.  
	Although they found electric dipole strengths commensurate with experimental observations, the calculated tunnel splitting as a function of O-O bond distance reached a minimum of approximately 16 GHz.  
	Such a lower limit is not seen in experiments and is inconsistent with the picture of TLS contributing to both low- and high-frequency noise (see section~\ref{Sec:Noise}).
	
	By a combination of ab-initio structures and single body Schr\"odinger equations, \textcite{Holder:PRL:2013} also investigated the role hydrogen plays in hydrogenated Al vacancies, 
	bulk hydrogen interstitial defects, and in a surface O-H rotor. 
	In this work, the structure was also computed using ab-initio methods to determine defect formation energy and stability.  
	The potential landscape seen by the rotor can then be extracted from the minimum energy pathway using the nudged elastic band method~\cite{Henkelman:JCP:2000}.
	In particular, they showed that the formation of hydrogenated Al vacancies is energetically favourable and that these defects form threefold degenerate rotors with tunnel splittings in the MHz to GHz range.  
	These defects also displayed electric dipole moments of approximately $0.6$~e{\AA}, in the range typically observed in experiments.
	
	Using empirical potentials, \textcite{DuBois:PRL:2013} solved the Schr\"odinger equation for the position of the oxygen atom - in analogy with oxygen interstitial defects in crystalline silicon and germanium.  
	Varying the position of surrounding aluminium atoms allows one to test various double-well configurations and compute the splittings to high precision using conventional finite-element techniques.  
	Although one can find many atomic configurations that show the correct range of splittings and charge dipoles, this still leaves the question of which atomic configuration is the correct one largely unanswered.  
	There are also significant differences in energy scale depending on whether a one-, two- or three-dimensional model is employed~\cite{DuBois:PRL:2013, DuBois:NJP:2015, DuBois:2015}.  
	In principle, both these limitations can be addressed by taking realistic atomic positions from ab-initio molecular dynamics simulations~\cite{DuBois:MS:2015}.  
	However, at stoichiometries and densities which are representative of experimentally grown oxides, the resulting TLS have splittings in the range of terahertz and above.  
	This suggests that the molecular environment within a junction is too tightly constrained to permit appreciable delocalisation of a single oxygen atom~\cite{DuBois:2015}, 
	providing further weight to the argument that clusters of atoms are involved in forming the TLS~\cite{Reinisch:PRL:2005}.
	
	In order to explore TLS arising from more realistic atomic configurations for amorphous Al$_2$O$_3$, \textcite{Paz:PRB:2014} performed molecular dynamics simulations at `elevated' temperatures of 25K 
	and searched for bistable switching between atomic configurations.  
	The free energy profile was then extracted for these configurations and barrier tunneling and charge dipoles estimated.  
	This approach allowed the identification of general structural motifs that display bistability without having to presuppose any particular symmetry of the defect.  
	Several candidate configurations were found with charge dipoles of order 0.9~e{\AA} and estimated energy splittings of 70~-~170 GHz.  
	However, the statistics of defect identification was limited by both the amorphous nature of the structures and the complexity of the calculation.

\subsection{Tunneling electrons}
	Following a similar philosophy to the tunneling atom models, the idea that single electrons can tunnel between local minima is also a very clear concept.  
	Many of the earlier experiments on strongly coupled TLS where analysed in terms of such electron tunneling~\cite{Lutchyn:PRB:2008, Koch:PRL:2007}
	However, this interpretation somewhat fell out of favour as the small energies of TLSs were taken as corroboration of tunneling atoms because of their larger mass as compared to electrons. 
	To contribute appreciably to thermal properties at sub-Kelvin temperatures, a significant number of TLSs need to have energies of the order of gigahertz, 
	which was considered inconsistent with the typical eV energy scale for electrons in solids~\cite{Agarwal:PRB:2013}.  
	
	More recently, the tunneling of electrons has been re-addressed considering more sophisticated effective models. 
	\textcite{Agarwal:PRB:2013} showed that if an electron moving between two wells is dressed by a collective phononic state, this has the effect of renormalizing the effective TLS parameters. 
	This renormalization factor was estimated to be of order $\ee^{-10}\approx 4.5 \times 10^{-5}$ and 
	the resulting renormalised TLS parameters were found to be typical energy scales commensurate with experiments. 
	The decoherence channels and TLS-TLS interactions one would expect for this model were also estimated and are compatible with observations in superconducting circuits. 
	Lastly, a detailed experimental approach to test this model was suggested, using phononic band-gap engineering of a metal-oxide junction to effectively suppress the phonon dressing 
	and thus gap out the TLS in the microwave range. {Such an engineered device provides a pathway to benchmarking different models and comparing their predictive power, 
	even for amorphous devices with a distribution of parameters.}
	
	Another effective single electron model is given by localised metal-induced gap states (MIGS) at a metal/insulator interface. 
	In this case, disorder at the interface localises a substantial fraction of MIGS electrons. The TLS is then formed by the magnetic moment of this localised electronic state. 
	\textcite{Choi:PRL:2009} performed a tight-binding analysis of an exemplary metal/insulator interface and showed that the expected areal density 
	and resulting low-frequency noise spectral function are consistent with observed data on magnetic field noise in SQUIDs. 
	Although this model was originally presented in an effort to explain the localised magnetic moments measured on the metal-oxide surface of SQUIDs, 
	it may equally apply to Josephson junction based defects - forming either a localised charge or spin defect. 
	In this spirit, the model was subsequently used to analyse the results of SET measurements of strongly coupled TLS~\cite{Pourkabirian:PRL:2014} (see section~\ref{sec:qubitspec}). 
	At this stage, little is understood on how susceptible MIGS are to decoherence and therefore whether the model is consistent with the long coherence times of TLS observed in qubit and resonator experiments.

\subsection{Spins and magnetic impurities}
	Another natural model for TLS is given by the intrinsic spin of electrons or atomic constituents, which may generate fluctuating magnetic moments. 
	Magnetic noise in superconducting circuits has been studied extensively from both a theoretical and experimental point of view. 
	The magnetic noise observed e.g.\ in superconducting quantum interference devices (SQUIDs) typically shows a $1/f$ spectrum with an amplitude of $A_{1/\text{f}} \sim 1\mu\Phi_{0} \text{Hz}^{-\frac12}$, 
	limiting the sensitivity of a range of magnetometry and sensing applications as well as severely reducing coherence times of many superconducting qubit types~\cite{Koch:Jetp:1983, Wellstood:APL:1987}.
	A recent review by~\textcite{Paladino:RMP:2014} provides a complete overview on this topic, and here we will only summarise some more recent results to complete the picture.
	
	Experimental studies have shown that the magnetic noise is predominantly generated in the native oxide encasing the circuits 
	and initial measurements showed that it scales linearly with device dimensions~\cite{Sendelbach:PRL:2008, Anton:PRL:2013}.
	Assuming the noise to be generated by electron spins, densities of $\sim5\times10^{17}\text{m}^{-2}$ have been inferred from measurements in a variety of SQUIDs~\cite{Sendelbach:PRL:2008} as well as on other, 
	non-superconducting material surfaces~\cite{Bluhm:PRL:2009}. 
	
	Very recent work on spin states on the surface of metal-oxides has shown strong evidence of the role of oxygen and spin-spin interactions. 
	For example, Lee~\etal investigated surface vacancy states on Al$_2$O$_3$ and SiO$_2$ using density functional theory~\cite{Lee:PRL:2014, Adelstein:2014},  
	and concluded that dangling bond states on the surface can form paramagnetic localised magnetic moments and can at least partially explain the low frequency magnetic noise. 
	In following work, \textcite{Wang:PRL:2015} used density functional theory to study molecular oxygen adsorbed to the surface of Al$_2$O$_3$, estimating a magnetic dipole of $~1.8~\mu_B$. 
	Using Monte-Carlo simulations of a spin lattice, they show that the flux noise generated by such a model was consistent with that observed in SQUIDs. 
	This analysis corresponds well with (and was inspired by) recent experiments illustrating the role of molecular oxygen~\cite{Kumar:PRA:2016} and hydrogen~\cite{ Samkharadze:PRA:2016, deGraaf:PRL:2017,deGraaf:2017} 
	in generating magnetic flux noise. 
	It has been shown that careful surface treatments to remove either of them results in significantly reduced loss in resonator circuits. 
	It remains unclear what, if anything, of this analysis can explain the emergence of strong interaction between quantum circuits and TLS. 
	As yet, no experiments have demonstrated a signature of adsorbed magnetic moments while also demonstrating strong 
	{coherent} coupling of the same {frequency-resolved} entity to a qubit or resonator degree of freedom.
		
\subsection{Emergent models\label{Sec:Models:Emergent}}	
	There are various proposals of emergent TLS models, in which the underlying degrees of freedom do not specifically display TLS-like behaviour but interaction with other collective degrees of freedom results 
	in effective TLS behavior. 
	
	This general idea has also been studied in depth by the glass community, for instance assuming localised phonon modes resulting from anharmonic local potential wells (the `soft-phonon' or `soft-potential' model) 
	see e.g.~\cite{Karpov:SSC:1982, Ilin:JETP:1987,Buchenau:PRB:1991,Finkemeier:PRB:1998,Laird:PRL:1991,Laird:JPhysCM:1996,Trachenko:JPhysC:2000,Zurcher:PRE:1997,Vural:JNCS:2011,Lubchenko:MolPhys:2006}. 
	Although the STM works very well to explain the thermal response of glasses below temperatures of 1K, 
	distinct deviations from STM behaviour are observed between 1K and 10K, which could be explained in terms of such localised modes. 
	More details on this model and its predictions in glass physics can be found in chapter 9 of~\textcite{Esquinazi:2013}. 
	However, as this model is largely indistinguishable from the STM at temperatures below 1K, it has received little experimental attention in superconducting devices. 
	The rapid closing of the superconducting gap as system temperatures approach the critical temperature of aluminium ($T_c\approx1.2$ K) makes experimental tests in this regime extremely challenging.
	Here instead we focus on models proposed recently in the context of qubit and resonator experiments, or those in which metal oxides of interest to these experiments have been explicitly discussed.
	
	One class of emergent model assumes pairs of trapping levels in the oxide barrier coupled to the superconductor, where a Cooper-pair couples simultaneously to a pair of states. 
	This has been dubbed the Andreev-level fluctuator~\cite{Faoro:PRL:2005}.
	Population and depopulation of the trap levels provides the fluctuating charge coupling to the electric field in the dielectric oxide, and potentially also modifies the critical current of the Josephson junction~\cite{DeSousa:PRB:2009}.
	This mechanism can provide the correct frequency and temperature behaviour for the noise generated by TLS, but requires an unphysical high density of states for the electron traps~\cite{Faoro:PRL:2006}.
	To remedy those shortcomings, \textcite{Faoro:PRL:2006} suggest that strong on-site repulsion of trap levels could lead to Kondo-like resonances close to the Fermi level. 
	These resonances are characterized by a Kondo temperature $T_{K}$, and  the interplay between superconductivity and Kondo-physics determines the occupancy of the trap. 
	For $T_{K} \sim \Delta$ this mechanism leads to a high density of localized states at low energy, much larger than the original density of charge traps~\cite{Faoro:PRL:2006, Faoro:PRL:2008}. 
	
	Further, \textcite{Muller:PRB:2009} conjectured that the observed strongly coupled TLS in phase qubits are formed from superradiant Dicke states of interacting microscopic TLS, 
	providing an explanation for the occurence of strong coupling for  only a small number of TLS.
	In this model, the strongly coupled TLS would exhibit higher levels with a quasi-linear level structure, an observation that is incompatible with experiments probing the structure of those defects.
	
	Finally, also interactions between TLS can lead to strong modifications of their underlying properties and distributions. 
	\textcite{Coppersmith:PRL:1991} conjectured that the universality seen in the STM stems from very strong dipolar interactions between microscopic TLS. 
	In this case frustration of the interaction leads to some of the TLS being effectively decoupled from the rest of the ensemble and dynamically free. 
	The properties of these free TLS will be universal in a large range of parameters~\cite{Coppersmith:PRL:1991}.
	
	Similarly, \textcite{Schechter:PRB:2013} have proposed that much of the universal behaviour seen in glasses comes about due to the interaction between two classes of two-level defects. 
	These classes are distinguished by their local symmetries with respect to inversion. Defects which are inversion symmetric about their mid point have (relatively) low bias energy ($\varepsilon/k_B < 10K$) 
	and do not couple to lattice phonons to first order. 
	In contrast, defects which are not inversion symmetric have higher frequencies (exceeding those typically probed in qubit or resonator experiments) and respond more strongly to phonons. 
	This difference in phonon response leads to three different TLS-TLS interaction energy scales via acoustic dipolar interactions: At sufficiently low temperatures, 
	the inversion asymmetric TLS effectively `freeze-out' due to their mutual interaction. 
	Hereby, they form an effective irregular spin-lattice which imposes a disordered local strain field upon the nominally symmetric low-frequency TLS. 
	The interaction energy scales therefore lead to a hierarchy of different responses as a function of temperature and provide a plausible explanation for much of the universal behaviour attributed to TLS.
	This model was motivated and tested using disordered crystals~\cite{GaitaArino:PRL:2011, Churkin:PRB:2014, Nalbach:NJP:2017} 
	but direct applicability to amorphous metal-oxides is yet to be shown conclusively, although recent work suggests 
	that nonequilibruim absorption measurements provide a method for probing such interacting TLS models~\cite{Schechter:NJP:2018, Burin:PRB:2018}.
	\\

{
\textbf{Summary} - 
The large number of microscopic models for TLS proposed in the literature poses a major challenge when trying to identify clear candidates. 
Many of the models are hard to distinguish experimentally, as their signatures in the data are very similar.  
Experiments that probe several TLS properties simultaneously may be needed to finally unambiguously determine the microscopic origin of the TLS, as we will discuss in the following. 
}


\section{TLS interactions\label{Sec:Interactions}}

Here we introduce the basic ideas of how TLS can couple to the dynamical degrees of freedom of their environments, including superconducting circuits as well as each other. 
These ideas are closely tied to the possible microscopic models for the origin of TLS as already reviewed in Sec.~\ref{Sec:Models}, and careful analysis of these interactions may lead to final identification of the TLS' microscopic origin. 

\subsection{Interactions with quantum devices}

	TLS in close vicinity to quantum circuits can couple to their dynamics by three basic mechanisms, explained in detail below. 
	An important concept here is the strong coupling regime between two quantum system, which in this case can be reached 
	when the coupling strength $g$ between the host circuit and an individual TLS is much larger then the dissipation rates of both circuit and TLS, $ g > \Gamma_Q,\,\Gamma_\mathrm{TLS}$. 
	Here $\Gamma_Q$ and $\Gamma_\mathrm{TLS}$ are the decoherence rates of quantum circuit and TLS, respectively.
	The strong coupling regime allows one to directly access and manipulate the TLS quantum state using the circuit as a bus, enabling new types of experiments that help to understand the TLS origin
	(see Section~\ref{sec:quantumdynamics}).
	
	\textbf{Charge fluctuations - }In the first model, TLS are perceived as atomic-sized electric dipoles, 
	which couple to the oscillating electric fields $\mathbf{E}$ in capacitive circuit components and tunnel junction barriers. 
	The coupling can be described by
	\begin{align}
		H_{\mathrm{charge}} = \frac{\left( \hat q - \frac12\, q_\mathrm{TLS}\, \sigma_{z}^{(p)}\right)^{2}}{2C} \,,
	\end{align}
	where $\hat q$ is the dynamical charge on the circuit capacitor $C$ and $q_{\rm{TLS}}$ is the change in induced charge on the capacitor associated with a change in the state of the TLS.
	Equivalently, one can describe this situation as an electric dipole, formed by the TLS, coupled to an electric field induced by the charges on the circuit capacitor~\cite{Martin:PRL:2005, Martinis:PRL:2005}. 
	The coupling strength between circuit and TLS is then given by 
	$g = \mathbf p \cdot \mathbf{E}$ where $\mathbf p$ is the TLS' dipole moment 
	and $\mathbf E$ is the electric field at the TLS position. 
	
	Charge TLS residing within the tunnel barrier of Josephson junctions can be exposed to relatively high electric fields of up to several hundred $V/m$ and can therefore readily be in the strong coupling regime.
	Charged TLS residing on the circuit substrate or in amorphous surface oxides of electrodes will still couple to the stray electric fields induced 
	by the circuit dynamics, albeit more weakly, and are thus thought to lead mostly to dielectric loss and energy relaxation~\cite{Barends:PRL:2013}.
	
	\textbf{Critical current fluctuations - }In the second model, the two TLS states are associated with different transparencies of a Josephson junction tunnel barrier, corresponding to a change of the junction's critical current. 
	Since the rate of Cooper-pair tunnelling across the tunnel barrier falls off exponentially with distance, at rough interfaces the current is transported through a discrete set of conductance channels. 
	Thus, blockage of one channel by e.g., a displaced charge may have a large impact on the total conduction, 
	especially for small area junctions~\cite{Wakai:PRL:1987, Nugroho:APL:2013, VanHarlingen:PRB:2004} and may bring TLS into the strong coupling regime.
	A similar magnitude of critical current variation might be due to a strongly coupled Kondo impurity in the junction dielectric~\cite{Faoro:PRB:2007, Faoro:PRL:2008a}.
	All these microscopic mechanisms will lead to a coupling between the TLS and a superconducting circuit  
	through a modification of the Josephson energy in the Hamiltonian,
	\begin{align}
		H_{\rm{current}} = \frac12 \delta E_{j} \sigma_{z}^{(p)} \cos{\hat \varphi} \,,
	\end{align}
	where $\sigma_{z}^{(p)}$ indicates the state of the TLS (c.f. Eq.~\eqref{eq:HTLS}), and $\hat \varphi$ is the superconducting phase difference across the circuit's Josephson junction.
	The coupling strength $\delta E_{j} = \frac{\Phi_{0}}{2\pi} \delta I_{c}$ between qubit and TLS is here directly proportional to the variation $\delta I_{c}$ 
	of the critical current corresponding to the two states of the TLS. Here, $\Phi_{0} = h / 2e$ is the superconducting magnetic flux quantum.
	
	The critical current coupling was presumed to generate the $1/f$ noise observed in JJs and SQUIDS~\cite{Shnirman:PRL:2005, Constantin:PRL:2007} 
	and also limit the coherence time of qubits through the associated fluctuations of the qubit energy~\cite{Simmonds:PRL:2004}. 
	However, the majority of qubit experiments performed in the last decade are consistently explained by assuming 
	that TLS couple to the Josephson junction's electric field rather than modifying its critical current, as explained previously.
	However, with the recent realisation that surfaces of quantum devices host a very high density of fluctuating magnetic moments~\cite{Kumar:PRA:2016, deGraaf:PRL:2017}, 
	the community has largely focussed on fluctuating magnetic fields as the origin of the low frequency noise in quantum electronics. As such the critical current model as an explanation for the observed $1/f$ noise has somewhat fallen our of favour. 
		
	One notable experiment which found direct indication that TLS modify the critical current of a Josephson junction was reported by~\textcite{Zaretskey:PRB:2013}. 
	Here, the spectrum of a Cooper-pair box was observed to be twinned, displaying multiple parabolas shifted both in frequency and offset in gate charge. 
	These observations are consistent with a TLS that couples to both the voltage across the junctions and the critical current, 
	where the latter was found to fluctuate by a large relative value of 30-40\%. As such a result was only reported once so far, its significance remains however unclear.
	
	Finally, \textbf{magnetic impurities} on the surfaces of the quantum circuits may provide fluctuations of the magnetic field threading SQUID loops of the circuits. 
	Such loops are typically used in superconducting electronics as an effective means to make the Josephson energy $E_{J}$ tuneable by an applied magnetic field~\cite{Makhlin:N:1998}. 
	The coupling will be decribed by
	\begin{align}
		H_{\rm{magnetic}} = E_{j} \cos{\left( \frac{2\pi}{\Phi_{0}} \left( \Phi_{x} + \frac12 \Phi_{\rm{TLS}} \sigma_{z}^{(p)} \right) \right)} \cos{\hat\varphi} \,,
	\end{align}
	where $\Phi_{\rm{TLS}}$ is the change in magnetic flux in the SQUID loop due to a change in the state of the TLS, and $\Phi_{0}$ is the magnetic flux quantum.
	For small fluxes $\Phi_{\rm{TLS}} \ll \Phi_{0}$, the interaction manifests itself as effective fluctuations in Josephson energies, very similar to the effect of a fluctuating critical current~\cite{Cole:APL:2010}. 
	Magnetic impurities arising from adsorbed surface spins such as molecular oxygen and atomic hydrogen are thought to be mainly responsible for the low-frequency magnetic noise, 
	as their coupling to the circuit dynamics is in general weak~\cite{Wang:PRL:2015, Kumar:PRA:2016, deGraaf:PRL:2017, Samkharadze:PRA:2016}. 
	However, recent experiments~\cite{deGraaf:2017} have demonstrated a correlation between removal of surface spins via annealing 
	and a reduction in the dielectric noise of resonators, suggesting a tantalising link between magnetic and charge noise.
	\\
	
\textbf{Determining the type of interaction -}
	There are several possibilities of how to distinguish the type of interaction a given TLS has with their hosting device. 
	In general, each type of interaction discussed above leads to a different term in the Hamiltonian. 
	With enough control over individual Hamiltonian parameters, experiments can be designed to determine not just the strength but also the exact form of the interaction, 
	which in turn might allow one to learn the microscopic origin of the TLS under study. 
	\textcite{Cole:APL:2010} compared experimental data on two strongly coupled TLS in a phase qubit with a range of theoretical models. 
	Their analysis was able to place strong bounds on the parameters of some of the microscopic models in the literature, 
	specifically ruling out magnetic dipoles and severely restricting the Andreev level fluctuator hypotheses, 
	but was ultimately not able to pin down a specific interaction model.
	\textcite{Zhang:PRA:2011} proposed a similar experiment using strongly coupled TLS in a flux qubit, as present e.g. in the experiments of~\textcite{Lupascu:PRB:2009}. 
	Here the different symmetries of the flux qubit Hamiltonian would make a spectroscopy experiment sensitive to different degrees of freedom than for the phase qubit used in~\textcite{Cole:APL:2010}, 
	allowing one to overcome the constraints of the earlier experiments and further constricting the microscopic models.
	
	Alternatively, in the case where the coupling between TLS and devices is not limited to a single type of interaction,
	one can test for cross-correlations between noise fluctuations in different Hamiltonian parameters to determine the type of coupling. 
	This method was first proposed for fluctuations 	in bias charge and critical current of the Josephson junction in a phase qubit~\cite{Faoro:PRB:2010}, but so far not implemented in experiments. 
	\\

{
\textbf{Strength of the interaction -}
	When talking about interactions between TLS and quantum circuits, an important concept is again given by the strong coupling regime, 
	i.e. when the strength of interaction $g$ is larger than the individual decoherence rates of both circuit and TLS, $g \gg \Gamma_{Q}, \Gamma_{TLS}$.
	In this case, the TLS is coherently coupled, allowing one to manipulate its state and probe its dynamics directly, as will be discussed in more detail in Sec.~\ref{sec:TLSinqubits}.
	In the opposite case of weak coupling, $g\ll\Gamma_{Q},\Gamma_{TLS}$, the interaction between circuit and TLS can in general be treated perturbatively and the effect of the TLS 
	will be to provide an effective noise spectral function to the circuit~\cite{Shnirman:PRL:2005, Muller:PRB:2009}. 
	This applies equally to individual as well as ensembles of TLS and is most easily probed with resonators (Sec.~\ref{sec:resonators}) although also applicable to qubits (Sec.~\ref{sec:DielectricLoss})
	}

\subsection{Interactions with their dissipative environment}
	
	Apart from interaction with the dynamics of the hosting device, TLS will almost always show dissipative dynamics, 
	characterized by incoherent state switching and fluctuations in TLS energy~\cite{Wuerger:1997}.
	Generally the time evolution of the TLS density matrix $\rho$ can be described by a master equation of the Lindblad form
	\begin{align}
		\dot \rho =& -\ii\comm{H_{\rm{TLS}}}{\rho} \nn\\
			&+ \Gamma_{\downarrow} \diss{\sigma_{-}}\rho + \Gamma_{\uparrow} \diss{\sigma_{+}}\rho + \frac12\Gamma_{\varphi} \diss{\sigma_{z}}\rho \,,
	\end{align}
	where $H_{\rm{TLS}}$ describes the TLS' free evolution, $\Gamma_{\downarrow}$ is the rate of dissipative transitions from the TLS' excited to its ground state (relaxation), 
	$\Gamma_{\uparrow}$ the rate of transitions from the ground to the TLS' excited state, and $\Gamma_{\varphi}$ is the TLS' pure dephasing rate.
	Here, $\diss{\hat o}\rho = \hat o \rho \hat o\hc -\frac12 \left( \op o\hc \op o \rho + \rho \op o\hc \op o \right)$ is a Lindblad dissipator, 
	describing incoherent processes associated with the operator $\hat o$.
	The strong coupling regime between TLS and its hosting circuit is possible if the TLS decoherence rate $\Gamma_{2} = \frac12\Gamma_{1} + \Gamma_{\varphi}$ is smaller than its coupling strength $g$ to the circuit. 
	Additionally, the circuit's decoherence rate (defined analogously) has also to be smaller than $g$. 
	Here $\Gamma_{1} = \Gamma_{\downarrow} + \Gamma_{\uparrow}$ is the inverse TLS lifetime. 
		
	The canonical source of dissipation and decoherence for TLS is a coupling to phonon modes in their hosting material~\cite{Esquinazi:2013,Wuerger:1997,Nalbach:NJP:2017}. 
	The physical mechanism of this coupling is the variation in groundstate energy in each well of the TLS due to the variation of the surrounding potential structure through interactions with phonons and strain, 
	and the subsequent variation in the asymmetry energy $\varepsilon$ of the TLS~\cite{Jackle:ZPhys:1972}:
	\begin{align}
		\varepsilon = 2\,\mathbf{\gamma} \cdot \mathbf{S} + 2\,\mathbf{p} \cdot \mathbf{E} + \varepsilon_0.
		\label{eq:asymtuning}
	\end{align}
	{Here, $\mathbf{\gamma}$ is a tensor defining the TLS' coupling strength to the strain field $\mathbf{S}$, the second term accounts for the coupling of the TLS' electric dipole moment $\mathbf{p}$ 
	to the electric field $\mathbf{E}$, and $\varepsilon_0$ is an offset imposed by the TLS' local environment. 
	For the magnitude of the so-called deformation potential $|\mathbf{\gamma}|$, typical values of order $\approx1\,$eV are found with TLS ensembles from acoustic experiments in glasses~\cite{Jackle:ZPhys:1972}, 
	consistent with observations of strain-tuned individual TLS in the AlO$_x$ tunnel barriers of Josephson junctions~\cite{Grabovskij:Science:2012}.} 
	For charged TLS in piezoelectric substrates, the additional electric field component $\mathbf E$ associated with the lattice vibrations will lead to enhanced coupling to phonons~\cite{Ku:PRB:2005} and thus stronger TLS dissipation. 
	
	The notion that two-level systems may also interact with conduction electrons is based on observations that TLS in metallic glasses posses enhanced energy relaxation rates~\cite{Black:PRL:1979}. 
	In quantum circuits, charged TLS located in surface oxides of superconducting electrodes can still interact with BCS-quasiparticles originating from incomplete electron pairing.	
	This mechanism is similar to the one proposed to be responsible for qubit dissipation at elevated sample temperatures, 
	where quasiparticles that are tunneling across a JJ can absorb energy from the qubit~\cite{Catelani:PRB:2011}.
	The interaction of quasiparticles with single TLS in the tunnel barrier of a phase qubit was studied theoretically by \textcite{Zanker:IEES:2016} 
	with experiments performed by \textcite{Bilmes:arxiv:2016} as described in more detail in Sec.~\ref{sec:TLSinqubits}.

\subsection{TLS-TLS interactions \label{sec:TLSTLSInteractions}}

	Mutual interaction between TLS may occur by both elastic and electric dipole coupling when the defect separation does not exceed a few nanometres.
	{Although such interactions are neglected in the standard tunneling model, 
	they have previously been invoked to explain the linewidth broadening of ultrasonically excited TLS ensembles in glasses~\cite{Arnold:SSC17:1972} 
	and their slow fluctuation dynamics~\cite{Black:PRB:1977,Burin:1998}.}
	
	The first direct observation of two strongly interacting and coherent TLS was reported by~\textcite{Lisenfeld:NC:2015} from experiments on phase qubits. 
	Here, strain-tuning spectroscopy (see Sec.\ref{sec:strainspec}) was used to map out the TLSs' energy levels, and the results found to be consistent with a dipolar interaction between two individual TLS described by 
	\begin{align}
		H_{\rm{TLS-TLS}} = g_{\rm{TLS}}\, \sigma_{z}^{(p,1)} \sigma_{z}^{(p,2)}
	\end{align}
	where $\sigma_{z}^{(p,i)}$ describes the state of TLS $i$ in its position basis. In these experiments, the mutual coupling strength $g_{\rm{TLS}}$ was found to be a substantial fraction of the TLS level splitting. 
	Earlier experiments in the same group had already shown similar but weaker interactions in other TLS, making it evident that TLS-TLS interactions are not uncommon~\cite{Grabovskij:Science:2012}.
	
	Several groups have pointed out that allowing for TLS-TLS interactions provides self-consistent distributions for STM parameters 
	that are closer to experimental reality than the canonical ones~\cite{Faoro:PRB:2015, Schechter:PRB:2013} 
	and can explain recent results on fluctuations in superconducting resonators and qubits~\cite{Burnett:NatCom:2014, Mueller:PRB:2015} 
	and on TLS dephasing under the influence of static 	strain~\cite{Lisenfeld:SR:2016, Matityahu:PRB:2016}.
	
\begin{figure}
		\includegraphics{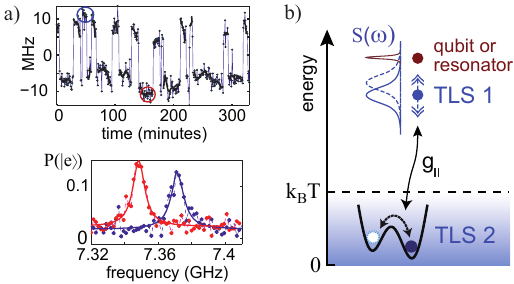}
		\caption{a) Top: Telegraphic switching and drifting of the resonance frequency of a TLS near 7.36 GHz~\cite{Lisenfeld:OwnData}. 
		Bottom: TLS resonances measured by direct microwave spectroscopy at the times indicated by blue and red circles in the top panel.
		b)  Illustration of the spectral diffusion mechanism. The resonance frequency of TLS 1 fluctuates due to its interaction with TLS 2, 
		which undergoes random thermal state switching since its energy is below $k_B T$. A qubit or resonator close to resonance with TLS 1 
		thus experiences a fluctuating environmental noise spectral density $S(\omega)$, affecting its energy relaxation rate and resonance frequency.}
		\label{fig:spectral_diffusion}
	\end{figure}

	Interactions between TLS are thought to be an important mechanism that gives rise to time-dependent fluctuations of quantum device parameters. 
	Here a high energy TLS interacts with one or multiple low-frequency TLS, whose excitation energy is below $k_{B} T$, such that they undergo random thermal transitions (i.e. fluctuators).
	In this case, the resonance frequency of the high-energy TLS may depend on the state of the fluctuator. If a TLS is coupled to one dominant thermal fluctuator, 
	its resonance frequency may display telegraphic fluctuations as shown in Fig.~\ref{fig:spectral_diffusion} a). 
	If more fluctuators are involved, continuous time-dependent drifts of the TLS resonance frequency may occur which is known as \emph{spectral diffusion}~\cite{Black:PRB:1977}. 
	
	This mechanism has the consequence that the noise spectral density which a TLS provides for a qubit or resonator at a given frequency fluctuates due to the time-dependent detuning between them, 
	as illustrated in Fig.~\ref{fig:spectral_diffusion} b)~\cite{Mueller:PRB:2015}. 
	Accordingly, qubits display time-dependent fluctuations in their energy relaxation rate as observed by~\textcite{Bertet:arxiv:2004, Paik:PRL:2011, OMalley:PRA:2015, Dial:SST:2016} and others, 
	who report fluctuations up to $\pm100$\% on a time scale of several hours. Most recently, these effects were observed using frequency tunable~\cite{Klimov:PRL:2018} and fixed frequency transmon qubits~\cite{Burnett:arXiv:2019, Schloer:arXiv:2019}, with the results well explained by the interacting TLS model. In microwave resonators, loss rate fluctuations up to 30\% were reported by~\textcite{Megrant:APL:2012} and attributed to the same mechanism.
	
	As a second consequence of this effect, the resonance frequency of a qubit or microwave resonator may fluctuate in time when its energy is dispersively shifted 
	by the coupling to a near-resonant TLS that undergoes spectral diffusion. 
	This causes qubit dephasing and poses a significant problem for envisioned superconducting quantum processors because qubits need to be re-calibrated at regular intervals.
	Such fluctuations were investigated by \textcite{Schloer:arXiv:2019} in a planar transmon qubit, revealing telegraphic switching of the qubit frequency on a time-scale of hours and associated changes in qubit decoherence rates. 
	In microwave resonators this mechanism causes resonance frequency fluctuations e.g. in the form of telegraphic noise as observed by~\textcite{Lindstrom:RSI:2011, Burnett:PRB:2013}, 
	as well as excessive phase noise at low frequencies and temperatures as studied by ~\textcite{Burnett:NatCom:2014}. 
	 
	\onlinecite{Meissner:PRB:2018} used mutual TLS interactions to observe the random state-switching of a TLF by monitoring the time-dependent frequency shifts of an interacting TLS, which itself was measured using a qubit. This technique allowed them to investigate the dynamics of TLF switching on time scales spanning milli-seconds to minutes.

	One other case where interactions between TLS are thought to be important is the origin of the low-frequency magnetic flux noise and specifically its frequency dependence. 
	The frequency dependence of the low-frequency flux noise is found to be $\sim 1/f^{\alpha}$ extending up to GHz frequencies~\cite{Slichter:PRL:2012, Quintana:PRL:2017}. 
	Here the exponent $\alpha$ is typically of order one and has been shown to depend on temperature~\cite{Wellstood:APL:1987}, 
	and its value has a strong influence on decoherence times of flux-sensitive qubits~\cite{Anton:PRB:2012}.
	Various models have been suggested to reproduce the low-frequency spectrum, 
	most of which involve interacting magnetic moments, with a variety of types of interactions under investigation~\cite{Atalaya:PRB:2014, Kechedzhi:2011, Faoro:PRL:2008, Carruzzo:PRB:1994, Chen:PRL:2010}.
	No clear consensus has been reached so far on the exact origin of the spectral signatures.
\\
	
{	
\textbf{Summary} - 
The fact that TLS interact not only with their hosting devices, but also with each other as well as their own environment is what makes them ultimately detrimental to the operation of superconducting devices. 
At the same time the interplay of interactions gives us an opportunity to unambiguously determine the microscopic origin of TLS by combining signatures from several different channels into a single experiment. 
In the following we will review the experimental progress so far towards this goal.
}


\section{Experiments with quantum bits\label{sec:TLSinqubits}}

\subsection{TLS microwave spectroscopy\label{sec:qubitspec}}

\begin{figure}
		\includegraphics{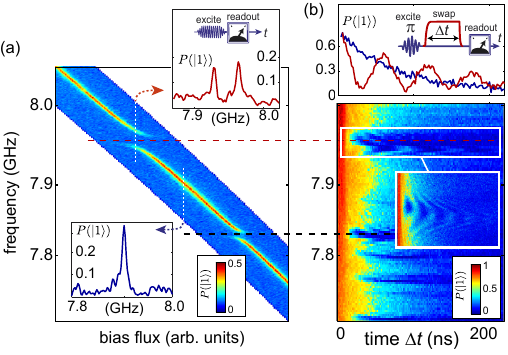}
		\caption{Qubit spectroscopy in (a) the frequency-domain and (b) the time-domain. 
			In (a), the qubit's excited state population $P(\ket{1})$ (color-coded) is measured after application of a long microwave pulse of varying frequency (insets). 
			Resonance with individual TLS (dashed horizontal lines) gives rise to split resonance peaks, so-called avoided level crossings. 
			(b) In the so-called ``swap-spectroscopy'' experiment, the qubit is first prepared in its excited state by a microwave $\pi$-pulse 
			and then tuned for a time $\Delta t$ to a varying probe frequency (see sequence in inset of upper panel). 
			While the isolated qubit shows pure exponential decay due to energy relaxation alone (blue line), the qubit tuned into resonance with a strongly coupled TLS displays additional oscillations (red line) 
			which reflect the redistribution of energy among the two systems due to quantum-state swapping~\cite{Lisenfeld:OwnData}.
		}
		\label{fig:qubit_spectroscopy}
	\end{figure}

	The development of superconducting qubits has provided significantly enhanced opportunities to investigate material defects because they can be used to detect individual TLS 
	and even allow one to control and observe their quantum state dynamics. The first signatures of strong interaction between qubits and single defects were found in microwave spectroscopy experiments. 
	Here, the qubit's excitation energy is varied in a range of a few GHz (typically by an applied magnetic field) and tracked by probing its population in response to application of long microwave pulses of varying frequency. 
	If the qubit is tuned into resonance with a strongly coupled TLS, the signal changes from a simple Lorentzian to a split peak due to the lifted degeneracy in the coherently coupled system. 
	See Fig.~\ref{fig:qubit_spectroscopy} (a) for an example of such measurements. 
	These characteristic avoided level- or anti-crossings were revealed in pioneering experiments on superconducting phase qubits performed in the group of J.M. Martinis~\cite{Simmonds:PRL:2004}, 
	whose observation that the distribution of anti-crossings changed once a sample was cycled to room temperature readily indicated the microscopic origin of the underlying TLS.
	Soon after, spurious resonances due to TLS were also observed in spectroscopy of flux qubits~\cite{Bertet:arxiv:2004, Plourde:PRB:2005} and in the so-called Quantronium, 
	which is a type of charge qubit consisting of a Cooper-pair box that is shunted by a large Josephson junction~\cite{Ithier:PRB:2005}.
	
	\textcite{Martinis:PRL:2005} also recognized that the major source of energy relaxation in first generation phase qubits was due to TLS-induced dielectric loss occuring in the junction barrier and its surrounding insulation layer. 
	This was tested by measurements on microwave resonators and qubits fabricated from different materials (AlO$_x$ and SiN$_x$) and confirmed by the relations found between qubit decoherence and dielectric loss tangents. 
	Moreover, an equation for the density of avoided level crossings observed in spectroscopy was derived from the standard tunneling model which reads
	\begin{equation}
		\frac{d^2 N}{dEdg} = \sigma A \frac{\sqrt{1-g^2/g^2_\mathrm{max}}}{2g},
		\label{eq:tlsnumber}
	\end{equation}
	where $E$ is the TLS energy, $g$ is the coupling strength between qubit and TLS, $g_\mathrm{max}$ is a maximal coupling strength determined by the largest observed TLS dipole moment, 
	$A$ is the junction area and $\sigma$ is the material-specific defect density. 
	Fits of this equation to the integrated number of observed splitting sizes show good agreement with experiments and provide a robust way to estimate the defect density $\sigma$. 
	The STM also agrees with the measured distribution of coupling strengths $g$ as verified by \textcite{Palomaki:PRB:2010} in a current-biased DC-SQUID.
	For large ($\approx 1 \mu m^2$) Al/AlOx junctions, typical TLS densities per frequency interval and junction area are found as 
	$\sigma \approx $ 0.4 - 0.5 (GHz $\mu m^2$)$^{-1}$~\cite{Martinis:PRL:2005,Simmonds:QIP:2009,Stoutimore:APL:2012}.
	 Assuming a typical tunnel barrier thickness of 2-3 nm, these measurements correspond to defect densities of $\sigma \approx 10^2 / (\mu m^3\:$GHz$)$ 
	 with maximal observed dipole moments of $p_\mathrm{max} \approx 6 - 8$ Debye. 
	 Slightly higher densities of $\sigma \approx 2.4 / (\mu m^2\:$GHz$)$ were reported for Al/AlO$_\mathrm{x}$ junctions by \textcite{Gunnarsson:SST:2013}, 
	 who used a custom process employing SiN$_\mathrm{x}$ insulation, 
	 and by \textcite{Hoskinson:PRL:2009} who investigated a qubit employing Nb/AlO$_\mathrm{x}$/Nb-trilayer junctions and found $\sigma \approx 2 / (\mu m^2\:$GHz$)$. 
	 For comparison, for silicon nitride Si$_3$N$_4$ which is known to have  significantly reduced dielectric loss, \textcite{Khalil:PRB:2014} extracted a TLS density of 
	 only $\sigma \approx 0.03 / (\mu m^2\:$GHz$)$  from measurements on lumped-element resonators. 
 	 	 	
	Spectroscopy on superconducting qubits can also be used to investigate the type of coupling between a qubit and TLS.
	As discussed in Sec.~\ref{Sec:Interactions}, the coupling can be longitudinal, where the energy of one system depends on the state of the other 
	(e.g. if the TLS affects the critical current of a Josephson junction), or transversal, where the TLS and its host circuit can exchange energy (e.g. a TLS coupled via its electric dipole moment).
	For example, \textcite{Lupascu:PRB:2009} observed the two-photon transition to the third excited state of a flux qubit-TLS system and found that TLS must be two-level or at least highly anharmonic systems 
	which are purely transversally coupled to the qubit. A similar analysis has been done with TLS in phase qubits~\cite{Bushev:PRB:2010, Du:2014}.
	Spectroscopic data on multi-photon transitions of TLS in a phase qubit by \textcite{Bushev:PRB:2010} were further analysed by~\textcite{Cole:APL:2010} in an effort to verify different microscopic TLS models.
	All of these works can be consistently explained by assuming that TLS couple to the capacitive elements of the quantum circuits via an electric dipole moment, 
	and some were even able to place strong bounds on alternative models~\cite{Cole:APL:2010}.
	
	Even if avoided level crossings are not observed in qubit spectroscopy, 
	their relatively strong coupling to TLS residing in other capacitive circuit components can still be detected by resonant enhancements of the qubit energy relaxation rate, 
	while the larger number of TLS located in regions of weaker electric fields contribute to a background relaxation rate that is independent or only weakly dependent on qubit frequency~\cite{Shnirman:PRL:2005}. 
	\textcite{Barends:PRL:2013} found qualitative agreement between such data obtained on so-called Xmon-qubits and Monte Carlo simulations of random TLS distributions, 
	in which defects were assumed to occur in a 3-nm thick oxide layer on the aluminium electrodes of the coplanar qubit capacitor at similar densities as verified for AlO$_x$ tunnel barriers. 
	
	When the qubit resonance frequency is being tuned quickly, Landau-Zener transitions may occur when the qubit is swept through resonance with strongly coupled TLS. 
	This results in a reduction of the readout fidelity of phase qubits~\cite{Cooper:PRL:2004} and in additional losses during qubit operations. 
	\\
	
	\textbf{TLS in charge qubits and single-electron transistors -}
	Experiments with charge qubits, Transmons and flux qubits only rarely observe avoided level crossings. 
	These devices employ very small Josephson junctions with typical areas of $0.01 - 0.1~ \mu m^2$, in contrast to phase qubits in which junctions have sizes of $1 - 10~ \mu m^2$. 
	Moreover, TLS densities in submicron-sized junctions were reported to be even lower than expected from the statistical scaling with junction area according to Eq.~(\ref{eq:tlsnumber}), 
	giving rise to speculation about self-annealing effects~\cite{Martinis:arxiv:2014} and the role of reduced film stress in smaller geometries~\cite{Oliver:2013}. 
	\textcite{Schreier:PRB:2008} tested in total ten Transmon samples and ten charge qubits and found avoided level crossings with splitting size exceeding 4 MHz in three of them, 
	roughly estimating the density of strongly coupled TLS to about $\sigma \approx 4.4 /(\mu m^2$ GHz). 
	When comparing this number to previous results on phase qubits, note that the smallest resolvable splitting size depends on the qubit's coherence time and thus spectroscopic line width.

	In charge qubits and single-electron transistors (SETs), a static electric field can be applied across the Josephson junction barrier, which for electrically active TLS tunes their asymmetry energy $\varepsilon$. 
	This could be directly observed by \textcite{Kim:PRB:2008} in a cooper pair box qubit, where the TLS resonance frequencies were found to depend linearly on the applied static electric field as expected 
	for asymmetric TLS formed by electric dipoles. In experiments on an SET, \textcite{Pourkabirian:PRL:2014} 
	observed that its effective charge bias was subject to a logarithmic drift after a sudden voltage step was applied to the gate. 
	This can be interpreted as originating in the slow relaxation of TLS into their new ground state due to inversion of their asymmetry energy by the gate voltage step. 
	The same group also studied the temperature dependence of charge noise in an SET and showed that the environmental TLS were in stronger thermal contact with the  SET electrons
	than with the phonons in the substrate~\cite{Gustafsson:PRB:2013}. The low-frequency noise in SETs due to TLS was investigated also in earlier work by~\textcite{Zorin:PRB:1996}, 
	 who focussed on the correlations between fluctuations in two adjacent SETs, and found that the responsible fluctuating charges were located either in the substrate or in the dielectric covering the circuits.

\subsection{TLS strain-spectroscopy\label{sec:strainspec}}

\begin{figure}[t!]
		\includegraphics{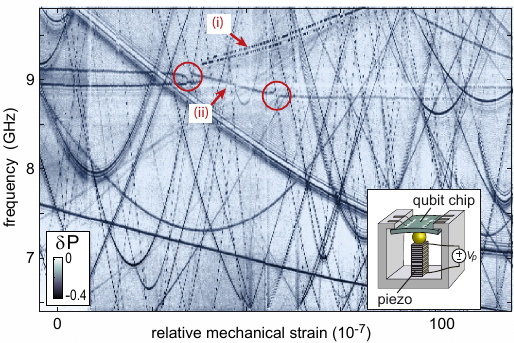}
		\caption{Resonances of TLS (dark traces) in dependence of the mechanical strain applied to a qubit chip. 
			$\delta P$ (color-coded) indicates the reduction in qubit population due to energy absorption from resonant TLS. 
			Mutual TLS interaction causes telegraphic switching of TLS resonance frequencies (i), non-hyperbolic traces (ii) and avoided level-crossings (circles). 
			Inset: Illustration of how the mechanical strain was controlled by bending the qubit chip with a piezo actuator.
		}
		\label{fig:tls_spectroscopy}
	\end{figure}

	The asymmetry energy $\varepsilon$ of a TLS depends linearly on the local electric field and mechanical strain as given by Eq.~(\ref{eq:asymtuning}).
	The latter effect provides a convenient way to tune TLS resonance frequencies in a given sample. To control the mechanical strain, \textcite{Grabovskij:Science:2012} 
	used a piezo actuator that slightly bent a chip containing a phase qubit (see inset of Fig.~\ref{fig:tls_spectroscopy}), 
	and spectroscopic measurements confirmed that TLS resonance frequencies indeed depend hyperbolically on the applied strain as expected from the standard tunneling model, 
	see Eq.~(\ref{eq:tls_energy}) and Eq.~(\ref{eq:asymtuning}). 
	From hyperbolic fits to strain-spectroscopy data, typical values of the TLS'  deformation potential of $\gamma \approx 0.1 - 1$ eV were extracted, which are consistent with measurements on bulk glasses~\cite{Phillips:1987}.
	
	Further strain-tuning experiments by the same group have provided expressive portraits of the TLS distribution as shown in Fig.~\ref{fig:tls_spectroscopy}~\cite{Lisenfeld:NC:2015}. 
	Here, TLS that were strain-tuned into resonance with the qubit were detected by their enhancement of the qubit relaxation rate. 
	Such data also reveal mutual TLS interaction in the form of avoided level crossings, non-hyperbolic traces, and telegraphic switching of TLS resonance frequencies.\\
	
	\subsection{Quantum dynamics of individual TLS\label{sec:quantumdynamics}}
	The strong interaction between a qubit and a TLS can be exploited to observe and manipulate the defect's quantum state. When a qubit is prepared in its excited state and tuned into resonance with a TLS, 
	the probability to find the excitation in the TLS will oscillate at a frequency that corresponds to the qubit-TLS coupling strength. 
	An example of this so-called \textbf{quantum state swapping} is shown in the upper panel of Figure~\ref{fig:qubit_spectroscopy} (b). 
	The strong coherent coupling to individual TLS is revealed by the characteristic ``Chevron''-type pattern as shown in the lower panel of Fig.~\ref{fig:qubit_spectroscopy} (b), 
	displaying the oscillatory redistribution of energy in the system. By setting the interaction time to half the inverse coupling strength, the quantum states of TLS and qubit are exactly swapped. 
	A TLS can thus be prepared in an arbitrary quantum state, and likewise the TLS' state can be read out by swapping it with the qubit's state where it becomes accessible for measurement. 
	This technique in principle allows one to use a TLS as a logical qubit as proposed by ~\textcite{Zagoskin:PRL:2006}, where the hosting superconducting qubit would merely be used for TLS manipulation and readout.
	
	Resonant swapping of quantum states between a phase qubit and a TLS has first been observed in the time domain by \textcite{Cooper:PRL:2004}. 
	Subsequently, \textcite{Neeley:NaturePhysics:2008} demonstrated the operation of a TLS as a quantum memory by storing the qubit state in the defect and recovering it after a waiting time. 
	Although contemporary qubit circuits show much longer coherence times than the randomly occurring defects considered here, certain kinds of TLS in nearly crystalline materials, 
	which presumably display a high degree of coherence,  may still become useful for quantum information processing applications.
	
	A phase qubit that is strongly coupled to a TLS displays beating Rabi oscillation when the system is resonantly driven~\cite{Ku:PRB:2005}. 
	This was studied as a function of drive amplitude~\cite{Sun:PRB:2012} and detuning \cite{Lisenfeld:PRB:2010}. In the latter work, it was found that a Raman-type transition exists in the detuned system 
	which allows one to directly manipulate the TLS' quantum state by resonant microwave driving while the qubit remains in its ground state. 
	This technique was used by \textcite{Lisenfeld:PRL:2010} to probe the temperature dependence of TLS energy relaxation and dephasing rates, 
	which at elevated temperatures were found to exceed the rates due to TLS-phonon coupling. 
	The responsible mechanism was identified by \textcite{Bilmes:arxiv:2016} to originate in the interaction of TLS with BCS quasiparticles, 
	where TLS couple to the evanescent electronic wave function that leaks from the junction electrodes into the tunnel barrier. 
	This work also showed that one may obtain information about the location of TLS across the tunnel barrier by injecting quasiparticles either into the junction's top or bottom electrodes, 
	which can provide clues about the fabrication step in which TLS predominantly emerge.
	\\
	
	
	\textbf{Entanglement} can emerge during the time-evolution of resonantly coupled quantum systems which share a single excitation. The entanglement between the state of a phase qubit that was tuned into resonance 
	with two TLS and a resonator was observed by \textcite{Simmonds:QIP:2009}, who found the dynamics of the qubits' state population to be consistent 
	with the emergence of a four-particle entangled system. \textcite{Grabovskij:NJP:2011} used a phase qubit to mediate entanglement between two TLS by tuning the qubit subsequently into the
	TLS resonances and performing a partial swap operation on each. A similar experiment probed the decay of an entangled state between a TLS and a resonator~\cite{Kemp:PRB:2011} and again found good agreement with theory. 
	The time evolution of different entangled states in a resonantly coupled qubit-TLS system was observed by~\textcite{Sun:PRB:2012}, 
	who also studied the emergence of tripartite entanglement via partial Landau-Zener transitions that occur when an excited phase qubit is swept through the resonances of two strongly coupled TLS.
	\\
	
	\textbf{Measurements of TLS decoherence times -}
	The ability to control and observe the quantum state dynamics of individual TLS provides a way to investigate the TLS' interaction with their local environment. 
	By monitoring the dependence of TLS decoherence rates on their strain-tuned asymmetry energy, \textcite{Lisenfeld:SR:2016} 
	found evidence that TLS phase coherence is limited by their interaction with thermally fluctuating TLS in their direct vicinity.  
	In addition, strain-independent maxima observed in TLS' energy relaxation rates at certain frequencies were attributed to the coupling of TLS to geometry-specific phonon modes in the Josephson junction.
	
	Swap spectroscopy was used by ~\textcite{Shalibo:PRL:2010} to obtain the statistics of the coupling strengths and coherence times of TLS in an AlOx Josephson junction tunnel barrier. 
	They observed TLS lifetimes $T_1$ between 12 ns and 6 $\mu$s which were on average anti-correlated with the coupling strengths to the qubit. 
	Their findings are consistent with the scaling of the radiative loss rate due to phonons with the defect's dipole size, as expected from the STM.

\subsection{Dielectric loss and participation ratio \label{sec:DielectricLoss}}

	When TLS are coupled through an electric dipole moment to oscillating electric fields within their host device, they can resonantly absorb energy and give rise to dielectric loss. 
	When trying to distinguish and quantify losses from different parts of a circuit, a useful concept is the participation ratio. 
	It specifies the fraction of the device's total energy that is contained in the lossy component or material. 
	The sum of all losses constitutes a limit for the circuit's total energy relaxation time $T_1$, which can also be described as an internal quality factor $T_1= Q_{\text{int}}/\omega$ \
	with the circuit's resonance frequency $\omega$. Employing the concept of the participation ration, we can write this as~\cite{Wang:APL:2015}
	\begin{equation}
		\frac{1}{T_1} = \frac{\omega}{Q_{\text{int}}} = \omega \sum_i \frac{p_i}{Q_i} + \Gamma_0 \,.
		\label{eq:partRatioT1}
	\end{equation}
	Here, $p_i$ is the participation ratio of the lossy component labelled $i$,  which has an internal quality factor $Q_i$,
	and $\Gamma_0$ is an additional dissipation rate accounting for non-dielectric losses. 	
	{According to the standard tunneling model, TLS that are themselves interacting with environmental electric fields and phonons cause dielectric loss rates of~\cite{Phillips:1987}
	\begin{align}
			\frac{1}{Q_\mathrm{i,el}} = \frac{\pi |\mathbf{p}|^2 D_0 }{3 \epsilon_i} \,,
			\ \ \mathrm{and} \ \ 
			\frac{1}{Q_\mathrm{i,ph}} = \frac{\pi |\mathbf{\gamma}|^2 D_0 }{2 \rho v^2},
			\label{eq:Qdef}
	\end{align}
	respectively. Here, $D_0$ is the (constant) TLS density of states, $\epsilon_i$ the permittivity of component material $i$, $\rho$ is the material density, 
	$v$ the sound velocity, and $\mathbf{p}$ and $\mathbf{\gamma}$ are the TLS' electric and elastic dipole moments, respectively.}

	Since the energy stored in a capacitive component scales with its capacitance $C_i$ and the voltage $U_i$ as $E_{i} = C_i 
	U_i^2/2$, 
	the participation ratio increases with the square of the electric field strength $\epsilon_i |\mathbf{E}_i(\mathbf{r})|^2$ integrated over the volume $V_i$ of the lossy component,
	\begin{equation}
		p_i = \int_{V_i} \frac{\epsilon_i}{2} |\mathbf{E}_i(\mathbf{r})|^2 / E_\mathrm{tot} \: d\mathbf{r}   \,,
		\label{eq:participation_ratio_calc}
	\end{equation}
	with $E_\mathrm{tot}$ the total electric field energy in the entire space. 
	Due to the high field concentration near tunnel junction electrodes and at edges of interdigitated capacitors, even thin dissipative layers in these regions may have a large impact on the device performance.\\
	
	A study of qubit energy relaxation rates and their dependence on the participation ratios of electrode surfaces by \textcite{Wang:APL:2015} 
	found conclusive evidence that dielectric dissipation is a major limiting factor in state-of-the-art qubits. 
	This result holds both for 3D-Transmon qubits, which were tested by varying the geometry of capacitor electrodes, as well as for planar Transmons, 
	for both of which coherence times above 100 $\mu$s have been demonstrated~\cite{Rigetti:PRB:2012, Martinis:private, Dial:SST:2016}. 
	
	Calculations of the participation ratio can be used to provide information on which region or interface of a quantum circuit contributes most of dielectric loss.	
	\textcite{Wenner:APL:2011} found the substrate-vacuum (S-V) and metal-substrate (M-S) interfaces to be 100 times more lossy than the metal-vacuum (M-V) interface (see Fig.~\ref{fig:CPW} for an illustration). 
	This is in accordance to \textcite{Dial:SST:2016} and \textcite{Sandberg:APL:2012}, who found an order of magnitude smaller participation ratio for the M-V interface ($p\approx 0.1\cdot 10^{-3}$) 
	as compared to the S-V ($p\approx 1\cdot 10^{-3}$) and the dominating M-S ($p\approx 3\cdot 10^{-3}$) interfaces. 
	The small participation of the M-V interface has been attributed to the large mismatch of dielectric constants~\cite{Martinis:arxiv:2014}.
	While above studies assumed the permittivity of the interfacial layer to be $\epsilon_r \approx 10$, \textcite{Quintana:APL:2014} point out that the relative contribution of the different interfaces depends strongly on $\epsilon_r$. 
	At low $\epsilon_r \approx 2$, corresponding to the permittivity of copolymer resist employed in fabrication, 
	the M-S interface was shown to still participate particularly strong, while the M-V and S-V interfaces now contribute about equally. 
	Employing this knowledge, it has been shown that the influence of TLS in the substrate can be reduced significantly 
	by etching a trench into the gap region~\cite{Barends:APL:2010,Wenner:APL:2011, Calusine:APL:2018,Bruno:APL:2015}.  	
	\\

{		
\textbf{Summary} - 
TLS in qubits were first identified as a major obstacle on the way towards useful superconducting quantum bits. 
However the unprecedented degree of control that is possible with these circuits, e.g. directly controlling the state of individual TLS and using the qubits as probes for their properties, 
leaves us with the very real possibilities to learn all there is to know about these mysterious defects. 
However qubit experiments and fabrication are particularly challenging, and in the following we will review other routes {to studying these same questions around TLS origin and behaviour}. 
}


\section{Experiments with superconducting resonators\label{sec:resonators}}

Superconducting microwave resonators (for a review, see ~\textcite{Zmuidzinas:Review:2012}) have recently found new applications as readout devices for superconducting quantum bits~\cite{Blais:PRA:2004, Wallraff:nature:2004}, 
for qubit interconnections~\cite{Sillanpaa:nature:2007, Mariantoni:science:2011}, as quantum memories~\cite{Hofheinz:nature:2009}, as quantum-limited amplifiers~\cite{Bergeal:nature:2010}, and as detectors for single photons, 
so-called kinetic inductance detectors (KIDs) \cite{Day:nature:2003}.

\begin{figure}
	\includegraphics{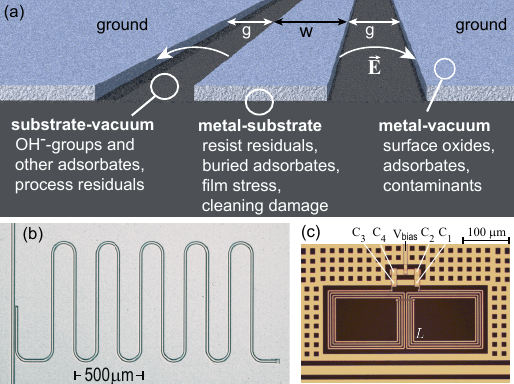}
	\caption{(a) Sketch of the cross-section through a coplanar transmission-line resonator, and an overview of mechanisms associated with TLS formation at different interfaces (circles). 
		(b) Photograph of a typical microwave resonator, having a total length of $\lambda/2 \approx$ 1 mm at a resonance frequency of 6 GHz, coupled to a transmission line. 
		(c) Lumped-element resonator which comprises a gradiometric inductor and four capacitors in a voltage-biased bridge~\cite{Sarabi:PRL:2016}.	
	}
	\label{fig:CPW}
\end{figure}

{Resonators are also playing an increasingly important role for the study of TLS in quantum devices, because they can be fabricated with the same technology and materials as quantum bits and are similarly characterized.}
Resonators for this purpose are typically shaped into a coplanar configuration as illustrated in Fig.~\ref{fig:CPW} a) and b), 
where a central conductor of a certain length is separated by gaps of a few $\mu$m size from the surrounding ground planes. 
Due to the presence of amorphous dielectric layers such as surface oxides, these devices will be susceptible to interactions with TLS, which in turn can be used to infer TLS properties. 
 
Alternatively, lumped-element resonators may be used, which comprise discrete planar inductors and capacitors. 
These usually employ meandering or coiled-up lines as inductors and interdigitated lines or overlapping films separated by a dielectric layer as capacitors, see Fig.~\ref{fig:CPW} c) for an example.
Lumped-element capacitors bring along the advantage that the electric field is mostly constrained to the dielectric volume of the capacitors and also homogeneous, 
greatly simplifying the estimation of the participation ratio of TLS-hosting dielectrics.
Also, the dielectric in a plate capacitor can be much better defined than the spontaneously emerging surface oxides which may be affected by contamination due to air exposure.	

Well-known effects that originate in coupling between microwave resonators and TLS are power-dependent resonator loss, a temperature-dependent resonance frequency shift, 
and excessive phase noise due to resonance frequency fluctuations and we summarise each of these in the following. 
In addition, these effects can be used to infer the densities of the TLS involved and resonator structures may allow one to directly manipulate TLS ensembles with applied electric fields.

\subsection{Power-dependent dielectric loss\label{sec:DielectricLoss2}}
	
	{The two-state character of TLS imposes a limit on their contribution to a resonator's loss rate: once a TLS was excited by the resonator field, 
	it has to first dissipate this energy and return to its ground state before a second photon can be absorbed. 
	When the circulating power $P_\mathrm{int}$ in the resonator exceeds a certain critical value $P_c$, 
	TLS are excited at an effective Rabi-frequency that exceeds their loss rates, $\Omega_R > 1/\sqrt{T_1 T_2}$. 
	This results in saturation of the TLS at a stationary excitation probability of close to $\nicefrac{1}{2}$ 
	with the consequence that the resonator's loss rate is reduced compared to the low-power limit.  
	Here, $\Omega_R = 2\mathbf{p\cdot E}\cdot \frac{\Delta_0}{E} / \hbar$ is the Rabi frequency for a TLS with dipole moment $\mathbf{p}$ in the resonant electric driving field $\mathbf{E}$, 
	with this TLS having an energy relaxation rate $1/T_1$, dephasing rate $1/T_2$, tunneling energy $\Delta_0$, and total energy $E$.}		
	The STM prediction for the resonator loss rate due to TLS saturation effects as a function of the circulating power $P_\mathrm{int}$ in the resonator is~\cite{Phillips:1987,Wang:APL:2009}
	\begin{align}
	    \frac{1}{Q_{\text{int}}} = \sum_{i} p_{i} \tan \delta_{i} \frac{\tanh \left (\frac{\hbar \omega_R}{2 k_B T} \right ) }{\sqrt{1+\left (\frac{P_\mathrm{int}}{P_c} \right )^\beta }}+\tan \delta_0,
	    \label{eq:tls_loss}
	\end{align}
	where $\tan \delta_{i}$ is the dielectric loss rate due to TLS in volume $i$ which has participation ratio $p_{i}$ (c.f Eq.~(\ref{eq:participation_ratio_calc})), $\tan \delta_0$ is a residual loss rate due to other mechanisms, 
	$\omega_R$ is the resonator's resonance frequency, and $T$ is the temperature. 	
	The sum goes over all lossy components that host TLS, effectively extending Eq.~\eqref{eq:partRatioT1} to saturation effects.
	The exponent $\beta$ is of order unity and for coplanar waveguide resonators is numerically estimated from the geometry to take into account the non-uniformity of the electric field distribution.
	
	The reduction of resonator loss $\propto 1/\sqrt{P_\mathrm{int}}$ in the few photon regime was observed in various experiments, e.g. by~\textcite{Lindstrom:PRB:2009,Pappas:IEEE:2011,Ramanayaka:arxiv:2015,Goetz:JAP:2016}.
	However, measurements on resonators with low intrinsic loss rates have shown a much weaker power dependence than the prediction of Eq.~(\ref{eq:tls_loss})~\cite{Burnett:SUST:2016,Macha:APL:2010,Sage:JAP:2011}.
	It has been suggested that this effect arises from spectral diffusion of strongly interacting TLS at interfaces and on surface oxides, 
	which causes TLS to drift through the resonator's resonance, effectively suppressing TLS saturation~\cite{Faoro:PRL:2012, Faoro:PRB:2015} 
	and leading to higher loss than predicted by Eq.~\eqref{eq:tls_loss}.
	Another possibility is the presence of two qualitatively different types of TLS ensembles with different critical saturation power $P_{c}$ as suggested by~\onlinecite{Schechter:PRB:2013}, 
	which would lead to a different saturation behaviour, as observed e.g. by~\onlinecite{Kirsh:PRM:2017}. 
		
	\textcite{Sage:JAP:2011} showed that it is possible to actively reduce the loss rate of a resonator by applying a strong microwave pump tone near the resonance frequency in order to saturate TLS. 
	This method is also known as \emph{hole burning}, which refers to the saturation-induced transparency enhancement first observed in materials which are doped with optically active bistable impurities 
	such as dye molecules~\cite{Esquinazi:2013}. 
	Recently similar experiments have probed the change of decay rate and frequency shift of the resonator when different parts of the TLS ensemble were saturated, 
	and found good agreement with the predictions from the STM for the spectral density of TLS even at very high frequencies~\cite{Capelle:2018}.

\subsection{TLS induced resonance frequency shift}
	
	The change of the resonance frequency of a resonator due to its coupling to a bath of TLS originates in the TLS' contribution to the dielectric constant $\epsilon$, 
	described by~\cite{Phillips:1987,Kumar:APL:2008,Gao:APL:2008}
	\begin{align}
		\frac{\Delta \epsilon}{\epsilon} = - \frac{2 D_{0}\, \mathbf{p^2}}{3 \epsilon}\left \{ \mathrm{Re} \left[ \Psi \left ( \frac{1}{2} + \frac{1}{2\pi i}\frac{h f_R}{k_B T}  \right ) \right] 
			- \log \left ( \frac{h f_R}{k_B T} \right ) \right \},
		\label{eq:dielectricconstant}
	\end{align}
	where $\Psi$ is the complex digamma function and $D_{0}$ the two-level density of states, c.f. Eq.~\eqref{eq:TLSDensity}. At low temperatures, 
	the resonant interaction with TLS in their ground state leads to an increased dielectric constant, 
	while at higher temperatures incoherent bath induced processes start to dominate and $\epsilon$ decreases again. 
	The eigenfrequency of resonators incorporating such dielectrics will scale accordingly as $\Delta f / f = - \frac{1}{2} \Delta \epsilon / \epsilon$.
	In contrast to loss, this frequency shift of the resonator also arises due to non-resonant TLS which are not saturated at high power levels, 
	providing a means to characterize the influence of TLS on a resonator also with measurements beyond the single-photon regime~\cite{Pappas:IEEE:2011}. 
	The non-monotonic temperature-dependent frequency shift was studied by \textcite{Gao:APL:2008} as a function of the center strip width $w$ of Nb coplanar resonators 
	(see Fig.~\ref{fig:CPW}). 
	In resonators where the electric field was more concentrated (for smaller strip widths and gaps), the frequency shift was more pronounced ($\propto 1/w$) as expected 
	for TLS that were distributed in a few nm-thick oxide layer on the surfaces of superconducting electrodes.  
	\textcite{Barends:APL:2008} showed that magnitude of the temperature-dependent resonance shift scales with the thickness of a SiO$_2$ layer deposited on top of NbTiN resonators, 
	clearly confirming the role of TLS in the amorphous capping layer.	
	
	Using the coupled symmetric and asymmetric modes of two overlapping coplanar waveguide resonators, \onlinecite{Kirsh:PRM:2017} investigated  
	the effect of the saturation of the TLSs by one mode on the quality factor and frequency shift of the second mode. 
	The availability of multiple modes in such a setup allows one to investigate a broader range of saturation effects, 
	i.e. the influence of strongly pumping the TLS in spectral vicinity of one mode on loss and frequency shift of the second mode.
	

	The influence of TLS on a resonators frequency also provides a novel pathway towards pinpointing the position of individual TLS and other impurities on surfaces.
	\textcite{Geaney:arxiv:2019} use measurements of frequency shifts of a microwave resonator on a quartz tuning fork 
	integrated in a scanning microscope setup to image the surface of a chip containing superconducting metal structures.
	Further improvements in thermal shielding and isolation of the noise background of the tuning fork resonator are necessary to achieve coherent coupling to individual TLS in that experiment, 
	which will then provide a clear way to identifying individual defects on surfaces, including their exact position and dielectric properties.

\subsection{Noise generated by TLS} 

	{Besides the so far discussed ensemble effects resulting in frequency shifts and resonator losses, 
	the coupling to single TLS may also result in a dressing of the resonator states and dispersive resonance shifts. 
	From the Jaynes-Cummings model it follows that in the weak coupling regime where the coupling strength $g$ is much smaller than the detuning $\Delta f$ between TLS and resonator, 
	$g \ll \Delta f$, the resonator experiences a dispersive resonance shift $\propto \pm g^2/ \Delta f$ depending on the TLS' state~\cite{Gao:APL:2007}. 
	Accordingly, spectral diffusion of near-resonant TLS can cause discrete resonance frequency fluctuations, 
	which in the ensemble limit translate into phase noise as discussed in the following section.}
	
	{Superconducting resonators are usually characterized by measuring the amplitude and phase of a resonant microwave pulse that is reflected on the resonator. 
	While fluctuations of the reflected amplitude concur with changes in the resonator's energy relaxation rate, phase noise is related to fluctuations of the resonance frequency. 
	Both effects can arise from spectral diffusion of a collection of near-resonant TLS, which can result in a time-dependent spectral density that determines energy relaxation, 
	as well as resonance frequency shifts due to the dressing of the resonator transition (see Fig.~\ref{fig:spectral_diffusion}).}
	
	
	Firm evidence that TLS are a source of resonator phase noise was obtained by \textcite{Gao:APL:2008b}, who found higher noise in resonators that had larger participation ratios at lossy interface regions.
	Further confirmation was provided with measurements as a function of the circulating power in the resonator. 
	The noise spectral density was shown to scale as $\propto P_\mathrm{int}^{-1/2}$, which indicates TLS saturation according to 
	Eq.~(\ref{eq:tls_loss})~\cite{Gao:APL:2007,Kumar:APL:2008,Barends:APL:2008,Lindstrom:PRB:2009}). 
	These findings are explained by \textcite{Gao:APL:2008b} using a semi-empirical model, where the dominant fluctuations are caused by TLS on the electrode surfaces which experience the strongest electric fields. 
	Similar conclusions were obtained by \textcite{Neill:APL:2013} studying the power-dependence of resonator loss and noise.
	
	Typically, fluctuations of the phase are found to dominate over amplitude noise by as much as 30 dB~\cite{Gao:APL:2007,Gao:APL:2011}. \textcite{Takei:PRB:2012} showed that this effect is due to squeezing of the noise quadratures by the nonlinearity of TLS coupled to the resonator, hereby enhancing the strength of phase fluctuations while amplitude noise is suppressed. 

	For increasing temperature, phase noise typically decreases in amplitude as $\propto T^{-1-\mu}$ in the single photon regime. 
	Here the exponent $\mu$ ranges from 0.2 to 0.7 \cite{Ramanayaka:arxiv:2015, Burnett:NatCom:2014}, and 
	is associated with the logarithmic temperature dependence of the spectral diffusion width $\Delta f(t,T)$, i.e. the spectral range over which a TLS diffuses over time~\cite{Burnett:SUST:2016, Black:PRB:1977}.
	These findings are consistent with a generalized tunneling model including interactions between high-frequency TLS and thermal fluctuators~\cite{Burnett:NatCom:2014,Faoro:PRB:2015}. 
	Within this model, at elevated temperatures spectral diffusion of near-resonant TLS plays an increasingly minor role because their transitions are already broadened thermally, 
	and the higher decoherence rates of TLS suppresses their interaction with thermal fluctuators. 
	Additionally, it is assumed that TLS at interfaces interact more strongly than TLS in the bulk, and that their interaction results in a suppression of the TLS density of states $\propto P_0 E^{\mu/2}$ 
	where $\mu$ is the same as the exponent in the temperature dependence of the noise. 
	An alternative explanation for these results was given by~\textcite{Burin:PRB:2015}, 
	who argue that mutual TLS interactions are less important at intermediate temperatures $T \geq 0.1 \, \mathrm{K}$  and assume that the TLS' spectral diffusion width is smaller than their relaxation rate. 
	This latter assumption was motivated by the early experiment of \textcite{Burnett:NatCom:2014} where a Nb resonator was capped with a normal-conducting Pt layer that was expected to enhance 
	TLS relaxation rates due to their interaction with quasiparticles.
	However, a later experiment by \textcite{Burnett:SUST:2016} showed consistency with the generalized tunneling model also for bare resonators.
	
	{Measurements of the frequency dependence of the noise power provide additional clues about the underlying physical mechanism. 
	The standard power-law model for noise predicts a scaling of the phase noise power spectral density $S_\varphi(f) \propto f^{-\beta}$ 
	where the integer values $\beta = 0, 1, 2, 3,$ and 4 are expected for white phase noise, 
	flicker phase noise, white frequency noise, flicker frequency noise, and random walk in frequency, respectively~\cite{Rubiola:arxiv:2005, Rubiola:book}. 
	Note that the power spectral density of the fractional frequency fluctuation $S_y(f)$ is related to that of phase noise by $S_y(f) = ({f^2}/{f_0^2}) S_\varphi(f)$.
	Early experiments, where the noise spectrum was probed at high circulating powers (with many photons in the resonator), obtained a frequency dependence close to 
	$S_y(f) \propto f^{-0.5}$~\cite{Gao:APL:2007,Baselmans:JLT:2008,Gao:APL:2008b,Barends:APL:2008}. 
	However, \textcite{Burnett:PRB:2013,Burnett:SUST:2016,Burnett:JPhys:2018} pointed out that those early results were likely influenced by instrument noise and short data acquisition duration, 
	and implemented an improved measurement setup based on a frequency-locked feedback loop~\cite{Lindstrom:RSI:2011} which resulted in clear evidence that the spectrum scales as $S_y(f) \propto f^{-1}$ (corresponding to $S_\varphi(f) \propto f^{-3}$),
	as expected for TLS-induced flicker frequency noise.}
	
	By covering resonators with various dielectrics and observing enhanced noise compared to bare resonators, 
	several experiments have directly confirmed the role of TLS hosting surface oxides as an origin of resonator noise~\cite{Barends:APL:2008,Burnett:PRB:2013}.
	Recently, \textcite{deGraaf:2017} observed a tenfold reduction in the magnitude of frequency fluctuations in NbN resonators after surface spins such as physisorbed atomic hydrogen were removed 
	by a thermal annealing treatment. In contrast, losses were reduced only weakly by spin desorption.
	This can be explained within the frame of the generalized tunneling model where surface spins take the role of the slowly fluctuating TLF that generate spectral diffusion 
	of the high-frequency atomic tunneling systems that are responsible for dielectric loss.
	
\subsection{TLS density measurements}
	The various effects TLS have on microwave resonators makes these devices useful tools to characterize defect densities in deposited materials, 
	and this will continue to be of importance in the search for low-loss materials for improved solid-state quantum devices. 
	 
	For example, \textcite{Bruno:PP:2012} measured the frequency dependence of the loss-rate of lumped-element resonators employing hydrogenated amorphous silicon (a-Si:H) dielectrics 
	and extract a relatively small loss tangent at 4.2 K of $\tan \delta = 2.5\cdot 10^{-5}$. Smaller TLS densities are observed in a-Si:H due to hydrogen atoms saturating dangling bonds and increasing the material density, 
	which curtails the atomic motional degree of freedom~\cite{Liu:PRL:2014,OConnell:APL:2008}. 
	In contrast, for thin  AlO$_x$ layers formed by plasma oxidation, \textcite{Deng:APL:2014} obtained the large value $\tan \delta \approx 1.6 \cdot 10^{-3}$ in agreement to other publications discussed here.
	
	Indications that the TLS density of states increases monotonically with energy were found by \textcite{Skacel:APL:2015}, 
	who analysed the frequency-dependent loss in lumped-element resonators made with amorphous SiO dielectrics. 
	This may stem from mutual TLS interactions, which are expected to decrease the density of states $n(E)$ at small energies due to the formation of an 
	Efross-Shklovskii type pseudogap~\cite{Burnett:NatCom:2014, Faoro:PRL:2012, Faoro:PRB:2015}. 
	However, more measurements and a systematic exploration of materials is necessary to confirm these findings.
		
	A robust technique to probe TLS densities in differently fabricated Josephson junctions was demonstrated by \textcite{Stoutimore:APL:2012}, 
	who devised a lumped-element resonator comprising a meandered inductor, an inter-digitated capacitor, and a Josephson tunnel junction in similarity to phase qubits.
	Strong coupling to individual TLS in the junction was measured spectroscopically by tuning the resonance frequency via an applied magnetic field and observing avoided level crossings, 
	which occurred at similar densities as found in phase qubit experiments. 

\subsection{Electric field tuning of TLS}
	TLS that possess an electric dipole moment respond to an applied electric DC field by variation of their asymmetry energy as expressed by Eq.~(\ref{eq:asymtuning}). 
	This was demonstrated for individual TLS with lumped-element resonators in which the capacitance was formed by four Al/SiN$_x$/Al parallel-plate capacitors in an electrical bridge design, 
	allowing one to apply an electric DC-field bias to the dielectric and hereby tune TLS resonance frequencies~\cite{Sarabi:PRL:2016}. Figure~\ref{fig:CPW}c) shows a photograph of the design.
	Using such a device, \textcite{Khalil:PRB:2014} observed the dependence of resonator loss on the sweep rate of the DC-field. 
	In the strong driving regime, where loss is typically reduced due to TLS saturation, enhanced loss occurs while the TLS are tuned in frequency by the sweeping the electric field. This is explained by that fact that TLS which are tuned through the resonator's transition frequency may absorb energy through Landau-Zener transitions,  while their saturation is suppressed due to their short interaction time with the driving field.	

	As demonstrated by \textcite{Matityahu:arxiv:2019}, similar experiments in a slightly different parameter regime may be effective to decrease dielectric loss from ensembles of TLS. 
	They consider electric field sweeps which periodically tune TLS through resonance with a resonator, in the limit where TLS coherence times are much longer than the sweep period. In this regime, subsequent Landau-Zener transitions of individual TLS typically interfere destructively such that the TLS' energy absorption rate is effectively reduced, resulting in lower dielectric loss as verified experimentally. This method can provide a path to actively decouple the TLS bath from a quantum circuit.
	
	To extract the distribution of TLS electric dipole moments $\mathbf p$, \textcite{Sarabi:PRL:2016} measured the hyperbolic signatures 
	of individual strongly coupled TLS in the resonator transmission 
	as a function of frequency and applied bias field. 
	Their experiments showed a broad maximum between one to three Debye and extending up to $\sim$ 8.3 Debye. 
	Moreover, avoided level crossings were observed in the transmission when strongly coupled TLS were tuned through the resonator's resonance.
	By fitting those to the Jaynes-Cummings model, the dipole moments and coherence times of these strongly coupled defects can be extracted.
	Similar results were obtained recently by \textcite{Brehm:2017}, who investigated TLS residing in an Al/AlOx/Al plate capacitor connected to the end of a transmission-line resonator. 
	By tuning TLS via an applied mechanical strain, resonances of individual strongly coupled TLS were detected and their dipole moments and energy relaxation rates were extracted. The strong resonator-TLS interaction of this system resulted in pronounced resonator frequency fluctuations due to spectral TLS diffusion.
		
	A random ensemble of TLS can even be used as a lasing (or strictly speaking masing) medium and coherently amplify the resonator excitation, as was demonstrated by~\textcite{Rosen:PRB:2016}. 
	In their experiment, TLS were first inverted into their excited states via Landau-Zener-transitions by sweeping them electrically through resonance with an applied microwave pump tone. 
	Afterwards, the excited TLS were tuned through resonance with a resonator, to which they transferred their energy by stimulated emission, generating the laser field. 
	\\

{	
\textbf{Summary} - 	
Resonators are used as tools for precision measurements in many fields, and the study of TLS in amorphous materials turns out to be one of them. 
Although they are mostly sensitive to effects from an ensemble of TLS, rather than individual ones, compared to qubits they allow for more rapid turn-around in experiment 
and thus systematic analysis of designs and fabrication parameters.  
In the quest to determine the microscopic origin of TLS, they are and will remain an indispensable part of the toolbox.
}


\section{TLS in other devices\label{sec:othersystems}}
There is a variety of other solid-state systems in which TLS appear to play a role. 
To highlight the connections to TLS in quantum circuits, in this section we briefly comment on examples where either the systems in which the TLS effects are observed displays quantum coherence, 
or where the effects of individual TLS can be measured or inferred. 
	
The coupling of TLS to strain and phonons gives rise to a mechanism of damping in nano- and micromechanical resonators such 
as suspended beams, cantilevers and membranes~\cite{Imboden:PRB:2009, Venkatesan:PRB:2010, Hoehne:PRB:2010, Suh:APL:2013, Faust:PRB:2014,Imboden:PhysRep:2014}, 
and was also shown to affect bulk acoustic resonators~\cite{Zolfagharkhani:PRB:2005, Seoanez:EPL:2007, Goryachev:APL:2010, Riviere:PRA:2011, Ahn:PRL:2003} and surface-acoustic wave resonators~\cite{Manenti:PRB:2016}. 
Here directly the coupling of TLS to phonons is responsible for opening an additional dissipation channel, similar in spirit to dielectric loss discussed in Sections~\ref{sec:DielectricLoss} and \ref{sec:DielectricLoss2}.
For a more complete overview of 
this field, see~\textcite{Aspelmeyer:NJP:2008} and references therein.

TLS even have a detrimental effect on the quality of optical devices such as lasers and atomic clocks when they reside in amorphous reflective coatings, where their mechanical fluctuation contributes to thermal noise. 
This was reported to be a limiting factor on the finesse of interferometers used in gravity wave detectors such as LIGO~\cite{Martin:CQG:2014, Trinastic:PRB:2016}.
In ion-trap experiments, TLS residing on trap electrodes were discussed as a source of so-called excess heating and electric-field noise~\cite{Brownnutt:RMP:2015}.

Recently, TLS were reported to be responsible for loss in collective electron-spin excitations known as magnons that were observed by coupling a piece of Yttrium-Iron-Garnet (YIG) to a cavity resonator
~\cite{Lachance:APE:2019, Pfirrmann:arxiv:2019}. In this case the coupling between magnon and TLS was assumed to occur via phonons.
In field effect transistors~(FETs), the switching of individual two-state defects was identified as the cause of telegraphic noise~\cite{Kirton:AIP:1989}. The same mechanism gives rise to blinking of fluorescent dye molecules (see \textcite{Orlov:JCP:2012} and references therein). 
Defect switching is also thought to be the origin of telegraphic conductance fluctuations observed in metallic nanocontacts at intermediate temperatures~\cite{Ralls:PRL:1988}. 
In such a system, the switching rate also varies with the applied mechanical strain as expected for atomic tunneling systems~\cite{Brouer:EPL:2001}.

A recent experiment by \textcite{TenorioPearl:NatMat:2017} studied FETs formed by gated 1D nanowires, 
whose charge distribution was deliberately disordered by defective capping layers of TiO$_2$ or Al$_2$O$_3$. 
Under application of a microwave drive, the transistor current displayed a large number of resonances having high quality factors $\approx 10^5$, 
whose resonance frequency changed once the sample was thermally cycled. 
Moreover, the transistor current showed oscillatory behaviour resembling Larmor precession and Rabi oscillation in response to pulsed resonant excitation, 
with decay times up to several tens of $\mu$s. 
The origin of these resonances was attributed to charged two-state defects which influence the conductivity of percolating current pathways in their vicinity. 
The observed resonance frequencies in the GHz range together with the sub-MHz linewidths indicate that these could very well be the same type of defects, 
discussed in Section~\ref{sec:quantumdynamics},
which are typically observed in superconducting qubits.



\section{Emergence of TLS in fabrication\label{sec:fabrication}}

It has become increasingly clear that TLS are associated with the formation of amorphous interface layers, disordered materials, and surface adsorbates.
In order to avoid and reduce loss from TLS in superconducting circuits, several strategies have been investigated, including:
\begin{itemize}
	\item remove lossy dielectrics wherever possible
	\item limit the device's coupling to TLS by employing circuit designs where electric fields are reduced
	\item employ less reactive superconducting materials to avoid amorphous surface oxides
	\item carefully optimize fabrication recipes to avoid TLS formation at interfaces
	\item fabricate electrode and tunnel junction barriers from crystalline materials.
\end{itemize}
Here we summarize some of the main results of recent studies on the effect of modified fabrication recipes and what they can tell us about origin and location of TLS.

\subsection{Removal of dielectrics}
	In conventional microcircuits, deposited dielectrics such as amorphous SiO$_2$ are frequently used as insulating layers for the realization of wiring cross-overs and on-chip plate capacitors. 
	In quantum circuits, these have to be avoided as they can contribute significantly to the total loss. Insulating cross-overs of coplanar resonators and transmission lines are therefore typically realized by so-called airbridges, 
	which are free-standing wire interconnects made by depositing a superconducting layer on a photoresist pedestal which is subsequently dissolved~\cite{Chen:APL:2014}. 
	Such airbridges are also required to equalize the ground plane potentials in coplanar resonators in order to avoid parasitic slot line modes. 
	Similarly, so-called vacuum gap capacitors have been realized by reactive ion-etching of the silicon dielectric that separates two overlapping aluminium electrodes, 
	achieving a reduction of capacitor loss by one to two orders of magnitude~\cite{Cicak:IEEE:2009,Cicak:APL:2010}. An often used alternative are planar capacitors in the form of interdigitated fingers, 
	for which part of the electric field is contained in vacuum. In this case, coupling to TLS on the substrate and electrode surface has to be avoided by limiting the electric field with an enhanced spacing between electrodes. 
	Experiments by \textcite{Sandberg:APL:2013, Gambetta:IEEE:2017} showed that energy relaxation in contemporary transmon qubits is dominated by capacitor loss when the finger spacing is below 20 to 30 $\mu$m. However, strong electric fields cannot be avoided near the qubit's tunnel junctions. In an effort to reduce substrate loss in this region, \textcite{Chu:APL:2016} etched away the silicon substrate to obtain freely suspended Josephson junctions and DC-SQUIDs. While this treatment resulted in longer qubit $T_1$ times, it also enhanced the level of flux noise, presumably because the exposed bottom edge contained a larger density of spins than the metal-substrate interface of reference samples.

\subsection{Superconducting materials and film deposition}
	To minimize the density of TLS in amorphous surface layers, devices were fabricated from superconducting materials such as rhenium and nitrides including NbN and TiN, 
	which are known to develop thinner oxide layers due to their weaker reactivity. 
	The loss rate of resonators fabricated from epitaxial rhenium was observed to be two to three times lower compared to sputtered aluminum~\cite{Wang:APL:2009}. 
	A study by \textcite{Sage:JAP:2011} compared resonators made from poly-crystalline Al, Nb, and TiN films, as well as epitaxial Al and Rh. They found TiN and Nb to have the lowest and largest losses, respectively. 
	In independent experiments, NbTiN was shown to exhibit much smaller loss~\cite{Barends:APL:2010} and less noise~\cite{Barends:APL:2010b} than Nb, Al and Ta. 
	Planar Transmon qubits made with TiN interdigitated capacitors on nitrided silicon substrates 
	show long coherence times up to 60 $\mu$s compared to similar devices made with Al capacitors that achieved $\approx 18 \mu$s~\cite{{Chang:APL:2013}}.
	
	However, in practice it is often difficult to attribute losses solely to the materials used, since their deposition and structuring typically involve different techniques and chemistry 
	with corresponding variations of surface morphology and residuals. 
	For example, sputtered aluminium results in rougher films and smaller resonator quality factor than Al that is evaporated via electron-beam or deposited by MBE~\cite{Megrant:APL:2012}. 
	Also the growth mode may have a large impact on losses as was shown by~\textcite{Vissers:APL:2010}, 
	who observed about a factor of 10 higher quality factors in resonators patterned from mostly (200)-TiN polycrystalline films compared to (111)-TiN. 
	A systematic comparison of TiN film properties as a function of sputtering parameters we presented by \textcite{Ohya:SST:2013}, showing that minimal losses of TiN resonators are associated with reduced film strain and, 
	surprisingly, increased oxygen content. Moreover, TiN is subject to ageing due to incorporation of contaminants once it is exposed to air, degrading device performance over time. 
	
	When fabricating tunnel barriers, \textcite{Tan:PRB:2005} found that the diffusion process in thermal oxidation of Al base electrodes may result in oxygen vacancies which bind a layer of chemisorbed O$_2^-$. 
	This in turn leads to excess junction noise and larger spread of barrier resistances over devices. These effects can be mitigated by the codeposition of Al and O$_2$ as shown by \textcite{Welander:APL:2010}. 
	They obtained ideal subgap resistances of amorphous AlO$_x$ barriers which were codeposited on epitaxial Nb/Al base electrodes, while thermal oxidation resulted in an excess shunt conductance.

\subsection{Dielectrics}
	Similar to superconducting nitrides, nitride dielectrics such as SiN$_x$ are typically less lossy than their oxide counterparts such as SiO$_x$~\cite{OConnell:APL:2008}. 
	The loss in amorphous dielectrics was shown to depend on the material density, indicating that the formation of atomic tunneling systems is inhibited in over-constrained materials. 
	For $a$-Si, this can be achieved by incorporating hydrogen ~\cite{OConnell:APL:2008,Pohl:RMP:2002} or by deposition at elevated temperatures which leads to more ordered and denser amorphous films~\cite{Liu:PRL:2014}. 
	In the latter work, it was shown that growth temperatures exceeding 350$^{\circ}$C result in TLS-free amorphous films exhibiting small loss rates $<2\cdot 10^{-7}$, 
	despite the fact that they contained no hydrogen but significant dangling bond densities as was verified by electron-spin-resonance measurements.
	
	More recent work has used transmission electron microscopy studies to correlate the structure of amorphous aluminium-oxide barriers with low temperature dielectric or tunnelling measurements~\cite{Zeng:JAP:2015, Zeng:SciRep:2016, Fritz:SciRep:2018}, demonstrating large variability depending on the growth conditions. In particularly, the influence of the aluminium metal contact morphology on the resulting aluminium-oxide film quality has been shown to be crucial.

\subsection{Substrates}
	Superconducting quantum circuits usually employ sapphire or high-resistivity silicon substrates. Compared to disordered interfaces, the loss rate of bulk crystalline substrates is several orders of magnitude smaller as was shown by
	\textcite{Creedon:APL:2011},who extracted loss tangents $\tan\delta < 10^{-8}$ for Sapphire (crystalline Al$_2$O$_3$) measuring the ring-down of dielectric whispering gallery mode resonators. 
	While the intrinsic loss of silicon substrates has not been well characterized at low temperatures, \cite{Martinis:arxiv:2014} reported that coplanar resonators fabricated on silicon perform slightly better than those on sapphire.
	Due to the minute contribution to the total dissipation of quantum circuits,  substrate loss is difficult to identify in experiments. 
	Moreover, direct comparison is complicated since different substrates may demand different clean-room processing techniques. 
	For example, \textcite{Gao:APL:2007} found smaller levels of phase noise for resonators produced on sapphire substrates compared to Si or Ge. 
	In contrast, \textcite{Sage:JAP:2011} reported little influence of the substrate when comparing different resonator and substrate materials, which may be explained by different processing steps employed in the former work. 
	No difference in resonator loss was found for silicon substrates that were capped with either a wet or dry oxide layer, 
	indicating that OH$^-$ groups were not a dominant source of TLS in these samples~\cite{Martinis:PRL:2005}. 
	Recently, \textcite{Dial:SST:2016} reported higher quality factors for Transmon qubits fabricated on sapphire substrates produced by the heat exchanger method 
	in comparison to the more common edge-defined film-fed grown sapphire. 
	{The lack of obvious correlations between such studies of substrate influence imply a very strong dependence on the particular fabrication procedure and its associated chemistry. 
	This in turn suggests that systematic (published) studies will be necessary if a general recipe is to be developed which can be implemented in a reliable way across different fabrication facilities.}
		
\subsection{Cleaning methods, chemical residuals and film structuring}
	To avoid defect formation due to adsorbates and residuals, proper cleaning of the substrate prior to material deposition turns out to be vital. 
	Thermally oxidized silicon substrates incorporate high densities of coordination defects at the Si/SiO$_2$ interface~\cite{Helms:RPP:1994}. 
	Therefore, cleaning Si substrates in hydrofluoric acid (HF) prior to film deposition in order to remove these native oxide and to terminate the surface with hydrogen 
	can significantly reduce resonator loss~\cite{Vissers:APL:2010,Goetz:JAP:2016}.
	\textcite{Megrant:APL:2012} showed that in order to fabricate resonators with quality factors exceeding $10^6$ from Al on sapphire it is crucial that the substrate is first cleaned, e.g. by a reactive oxygen plasma at 850$^{\circ}$C. 
	This is an indication that TLS are formed from adsorbed hydroxyl groups, which are capable of saturating the sapphire (0001) surface~\cite{Ahn:surf:1997} 
	and are stable enough to remain even after annealing at 1100$^{\circ}$C in UHV~\cite{Niu:surf:2000}.
	
	The cleaning method using a reactive oxygen plasma creates less substrate damage than ion-milling, resulting in smoother films and a higher quality interface. 
	\textcite{Quintana:APL:2014} obtained twice as large resonator quality factors and thinner disordered interfacial layers when the substrate was cleaned by weak
	in situ ion-mill (200 eV / 4 mA for 10 seconds) compared to stronger milling (400 eV / 20 mA for 3.5 minutes). This may indicate that the incorporation of Argon ions 
	or the (disordered) re-deposition of removed material is associated with TLS formation.	
		
	The importance of proper cleaning was further emphasized in a study of ageing effects in Josephson junctions by \textcite{Pop:JVAC:2012}. 
	Junction ageing is predominantly attributed to the presence of chemical contamination such as photoresist residues, which thermally diffuse into the junction barrier over time. 
	Pop~\etal showed that completely stable Al/AlO$_x$/Al junctions can be obtained if the substrate was initially thoroughly cleaned with an optimized reactive oxygen plasma. 
	This was further cross-checked by annealing junctions in the presence of a PMMA resist capping layer, which resulted in unstable junctions, presumably due to incorporation of hydroxides from the resist into the tunnel barrier. 
	This suggests that atomic diffusion of contaminants ({either from the substrate preparation or from the resist}) may significantly alter the material quality.
		
	The stability of  Al/AlO$_x$/Al tunnel junctions can also be improved by annealing finished junctions in a vacuum chamber at a temperature of 400$^{\circ}$C \cite{Koppinen:APL:2007}.  
	Moreover, it was shown that annealing also improves the tunnelling characteristics overall, as it results in increased subgap resistance and the additional disappearance of subgap resonances 
	which are due to resonant or inelastic tunneling at barrier impurities. 
	It was suggested that the elevated temperatures in the annealing process result in dissociation of aluminium hydrates which may have formed during thermal oxidation in the presence of water vapour~\cite{Gates:JAP:1984}. 
	The hereby released oxygen is expected to combine with the junction electrode material which further increases the tunnel barrier and reduces the critical current, in accordance with the previous results.
		
	Outgassing of PMMA photoresist masks during film deposition and their residuals due to incomplete development is suspected to degrade the quality of Nb resonators as reported by~\textcite{Chen:SST:2008}. 
	This effect may explain the (by a factor of 3) higher loss rates of resonators patterned in lift-off processes compared to etched resonators as reported by \textcite{Quintana:APL:2014}. 
	In a lift-off process, the substrate is covered with photoresist prior to metal deposition, and incomplete development may leave residual photoresist at the substrate-metal interface. 
	\textcite{Quintana:APL:2014} investigated these residuals by high-resolution transmission electron microscopy, revealing the presence of an 1.6 nm thick resist polymer layer showing a peak in carbon content, 
	followed by a 2 nm thick layer containing oxygen and AlO$_x$. The latter was presumably formed by a reaction of the deposited Al with resist or solvent residues and likely contained high TLS densities. 
	Moreover, it was shown that these residuals can be efficiently cleaned using a downstream oxygen ash descum procedure, 
	which employs neutral oxygen atoms to chemically remove developed photoresist from the heated substrate, 
	bringing the losses back to the lower levels found in the etched control resonators. Similar cleaning efficiency was found with an UV-ozone process that does not require substrate heating, 
	and is therefore preferable for Josephson junction fabrication.	
		
	\textcite{Sandberg:APL:2012} investigated how different etching methods affect the quality of TiN CPW resonators on Si substrates, 
	and found that Ar-ion mill patterning caused the formation of amorphous fence-like structures at the electrode etches due to re-deposition of Silicon. 
	{Due to their disordered structure, these fences likely contain large densities of TLS which will contribute significantly to the total loss because they are located in regions of higher electric field concentration.}
	Lowest loss was achieved using a fluorine-based reactive ion etch, presumably due to its higher edge rates. In contrast to the fluorine-based treatment, 
	the chlorine-based process was suspected to leave Cl salts on the surface and to implant boron ions, leading to more potential contaminants and higher dielectric loss.

\subsection{Epitaxial films}
	Soon after it was first recognized that qubit energy relaxation and appearance of avoided level crossings are associated with TLS in amorphous materials, 
	first attempts were made to realize completely crystalline tunnel junction barriers.
	\textcite{Simmonds:2004} compared the performance of Josephson junctions made with conventional ion-mill recipes to those from trilayer processes, 
	where the amorphous tunnel barrier is grown \insitu on a crystalline Al bottom electrode without breaking vacuum. 
	While DC transport characteristics such as residual subgap conductance were improved significantly for crystalline base electrodes, phase qubits made from these junctions showed no improvement. 
	However, the high-frequency loss relevant for qubit dissipation may have originated from the lossy SiO$_2$ dielectric used for junction insulation in all tested samples in that experiment.
	\textcite{Patel:APL:2013} fabricated crystalline shunt capacitors by etching a silicon-on-insulator substrate into a membrane which was covered by Al on both sides. 
	Phase qubits employing these shunt capacitors showed $T_1$ times twice as long as samples with amorphous dielectrics.
	
	Various problematic effects in the fabrication of epitaxial aluminium films were identified by \textcite{Richardson:SST:2016}. 
	These include corrosion of the Al sidewall due to resist developers, contamination by nanoparticle, and persistent photoresist residue - all of which could lead to an increased defect density.
	Josephson junctions with a crystalline Al$_2$O$_3$ tunnel barrier, grown on an epitaxial Re bottom electrode and capped by polycrystalline Al, were fabricated by \textcite{Simmonds:SST:2005}. 
	Phase qubits employing these junctions displayed a factor of five smaller number of avoided level crossings in spectroscopy, indicating the better crystallinity of the dielectric. 
	However the qubit's energy relaxation rate was not reduced compared to previous samples, presumably due to the presence of lossy SiO$_2$ wiring insulation in these early experiments~\cite{Oh:PRB:2006}. 
		
	Fully crystalline junctions were fabricated from epitaxial NbN/AlN/NbN trilayers deposited on MgO (100) substrates by \textcite{Nakamura:APL:2011}.
	Here special care was taken to grow the AlN tunnel barrier with a cubic crystal structure in order to avoid piezoelectricity which would lead to conversion of Josephson plasma oscillations into phonons. 
	Transmon qubits with these junctions showed T$_1$ times ranging between 250 and 450 ns. 
	This was significantly longer compared to $\sim$10 ns that was achieved with amorphous AlN tunnel barriers in early phase qubits~\cite{Martinis:QIP:2009}, 
	but also significantly shorter than state-of-the-art Transmons which employ shadow-evaporated junctions.
		
	Yet another approach was taken by \textcite{Kline:SST:2012}, who fabricated junctions with crystalline Sapphire (Al$_2$O$_3$) barriers grown on an epitaxial Re/Ti multilayer base electrode. 
	The multilayer was made by depositing 1.5 nm of Ti on 10 nm-thick layers of Re and repeating these steps 12 times. 
	This resulted in much smoother films (rms roughness of 0.6 nm) as compared to pure Re films which showed terrace structuring and higher rms roughness of 3.2 nm due to basal-plane twinning~\cite{Welander:JAP:2010}. 
	However, Junctions that are capped by Re show much smaller subgap resistance and poor performance when employed in a phase qubit as compared to junctions using Al top electrodes.
	\textcite{Weides:APL:2011} probed coherence in Transmon qubits which comprised (ReTi)$_{12}$/Al$_2$O$_3$/Al epitaxial junctions, and measured a small loss tangent of 6$\cdot 10^{-5}$ for the crystalline tunnel barrier. 
	Spectroscopy on these samples showed a small number of avoided level crossings due to strongly coupled TLS. 
	However observed energy relaxation times were still rather short, {although again this could simply reflect the less mature fabrication and design process compared to existing processes and designs}.

{
\textbf{Summary} - The results discussed in this section clearly indicate that TLS formation depends sensitively on the employed materials and techniques of film deposition, 
structuring, morphology, and substrate treatment. It still remains far from clear which combination of fabrication processes promises the greatest improvement in coherence and performance of quantum circuits. 
Nevertheless, it will be crucial that this question is tackled soon, given the potential benefits of a superconducting quantum processor and the importance of TLS induced noise and loss for its operation.
}

\section{Conclusions and outlook}
In this review, we have summarised the role TLS defects play in superconducting circuits, focusing on both the challenges and opportunities that they represent. 
Although TLS physics is a relatively old topic of study, superconducting circuits have provided an entirely new way of looking at this problem, and an even greater urgency to solve it.
Using superconducting circuits to probe TLS provides a range of new possibilities that were simply unavailable to the solid-state glass community over the previous decades. 
Probing an individual TLS, determining its characteristic properties including energy splitting, decoherence rates, and strain response, 
all as a function of applied fields, sample temperature or mechanical strain has now become achievable. 

Equally on the side of theory, there are now very specific targets and goals. 
The number of relevant materials is relatively small, the properties of interest are very specific and quantitative theory/experiment comparison is becoming the norm. 
We are now reaching the point where atomistic models can be compared directly to the (until recently) purely phenomenological parameters of the standard tunnelling model.
{Further theoretical advances have the potential to actually predict device parameters based on fabrication conditions, and significantly reduce the experimental turn around times.
}

To move forward will require ever closer collaboration between theory and experiment. 
New experimental designs and apparatus that allow measurement of several different parameters of a particular TLS need to be developed to be able to estimate these parameters without any unknown scale factors. 
Honest evaluation of the data with respect to all relevant theory models is required, without bias towards any particular model.
{Particularly promising here are efforts that determine the exact type of interactions between TLS and their hosting device, e.g. by exploiting the symmetries of the device Hamiltonian. 
Knowledge of the type and magnitude of the interactions, together with support from theory, will allow us to constrain the large zoo of theory models in the literature. 
Similarly, methods that probe the position of individual TLS or TLS ensembles, be it through systematic variation of participation ratios or through applied static or spatially dependent fields, have large potential 
to advance our knowledge of TLS origins when combined with systematic variations in device fabrication.
Finally, a largely unexplored frontier is the modification of the TLS environment, be it through phononic bandgap materials or quasiparticle generation or trapping. 
This again can give us valuable hints towards the microscopic origins of TLS, while at the same time already working towards minimizing their impact. 
}

Systematic studies of materials, fabrication methods and circuit design dependent performance will provide new insight and new solutions to this problem. 
{In particular, this should include making publicly available studies and catalogues of the fabrication processes used, tested and/or discarded. 
This is particularly important if the technology is to move from in-house recipes suitable for research publications to large scale device engineering.}
{The experiments needed to advance here are not quick and do not lend themselves to short format, single journal article studies.
Perseverance is necessary to prosper and make progress in this area, but the rewards can be worthwhile. 
}

TLS continue to be the Achilles heel of superconducting electronics but there is reason for hope. 
Ironically, using superconducting devices to probe TLS physics provides the very real possibility of solving one of the great mysteries of solid-state physics - the true microscopic origin of these ubiquitous defects.

\acknowledgements{
	We thank S.~Meissner, M.~Schechter, and A.~Seiler for useful comments and suggestions on the manuscript, and thank J. Burnett for valuable comments on TLS-induced phase noise.
	The authors would also like to acknowledge many useful conversations with the participants of the ``Tunneling two-level systems and superconducting qubits'' workshop 
	held in Sde Boker, September 2016 as well as the workshop ``Atomic tunneling Systems and fluctuating Spins interacting with superconducting Qubits'' at MPIPKS in Dresden, February 2019.
	We thank Lukas Gr\"unhaupt (Physikalisches Institut, Karlsruhe Institute of Technology) for providing Figs.~\ref{fig:QubitsAndTLS}b) and \ref{fig:QubitsAndTLS}c).
	JL acknowledges funding from Deutsche Forschungsgemeinschaft (DFG), project LI2446/1-1. 
	This work was supported by the Swiss National Science Foundation through NCCR QSIT and by the Australian Research Council under the Discovery and Centre of Excellence funding schemes (project numbers DP140100375, CE170100026 and CE110001013).
}

\pagebreak

\section*{Table of symbols}

\begin{table}[th!]
	\caption{\label{tab:SymbolsAll} Table of all symbols and their definitions.}
	\begin{tabular}{@{}ll}
		\hline
		Symbol		&	Definition\\
		\hline
		$\Delta_0$	&	Tunnelling energy between wells\\
		$\varepsilon$	&	Energy asymmetry between wells\\
		$E_{\pm}$		&	Eigenenergies of the STM Hamiltonian\\
		$d$			&	Distance between the wells in the STM\\
		$V$			&	Height of the barrier between the wells\\
		$P(\varepsilon, \Delta_{0})$	& 	Probability distribution of parameters\\
		$D(E)$		& 	Density of states\\
		$S(\omega)$	& 	Noise spectral function\\
		$g$			& 	Coupling strength\\
		$C$			& 	Capacitance\\
		$E_{j}$		& 	Josephson energy\\
		$\Phi_{0}$		&	Magnetic flux quantum\\ 
		$\Gamma$	&	Rate of an incoherent process\\
		$\Gamma_{1}$	&	Energy relaxation rate\\
		$\Gamma_{2}$	&	Total dephasing rate, including relaxation processes\\
		$\Gamma_{\varphi}$	&	Pure dephasing rate\\
		$\mathbf\gamma$	&	TLS strain coupling tensor\\
		$\mathbf S$	& 	Strain field tensor\\
		$\mathbf p$	& 	Dipole moment vector\\
		$\mathbf E$	&	Electric field vector\\
		$\sigma$		& 	Defect density\\
		$Q$			& 	Quality factor \\
		$T_{1}$		&	Energy relaxation time ($1/\Gamma_{1}$)\\
		$p_{i}$		& 	Participation ratio of material component $i$\\
		$\epsilon_{i}$ 	&	Dielectric constant of material component $i$\\
		$\tan{\delta}$ 	& 	Loss tangent\\
		$T$			& 	Temperature\\
		$P_{\text{int}}$	&	Circulating power in resonator\\
		$P_{\text c}$	&	Critical saturation power\\ 			
		\hline
	\end{tabular}
\end{table}

\bibliography{TLSreviewbibfile}

\end{document}